\documentclass[notitlepage,a4,10.5pt]{article}

\usepackage{graphicx}

\usepackage{amsmath}%
\usepackage{amsfonts}%
\usepackage{amssymb}%
\usepackage{mathrsfs}
\usepackage{amsthm}
\usepackage{float}

\usepackage{authblk}

\usepackage[left=2.5cm,right=2.5cm,top=3cm,bottom=3cm]{geometry} 

\usepackage{xcolor}

\newcommand{\mc}{\mathcal}
\newcommand{\cp}{\times}

\newcommand{\bol}{\boldsymbol}

\newcommand{\abs}[1]{\left\lvert{#1}\right\rvert}

\newcommand{\lr}[1]{\left({#1}\right)}

\newcommand{\mf}{\mathfrak}
\newcommand{\p}{\partial}

\newcommand{\ti}[1]{\textit{#1}}
\newcommand{\tb}[1]{\textbf{#1}}

\begin{document}

\title{Vorticity equation on surfaces with arbitrary topology}
\author[1]{Naoki Sato} \author[2]{Michio Yamada}
\affil[1]{Graduate School of Frontier Sciences, \protect\\ The University of Tokyo, Kashiwa, Chiba 277-8561, Japan \protect\\ Email: sato\_naoki@edu.k.u-tokyo.ac.jp}
\affil[2]{Research Institute for Mathematical Sciences, \protect\\ Kyoto University, Kyoto 606-8502, Japan
\protect \\ Email: yamada@kurims.kyoto-u.ac.jp}
\date{\today}
\setcounter{Maxaffil}{0}
\renewcommand\Affilfont{\itshape\small}

    \maketitle
    \begin{abstract}
		We derive the vorticity equation for an incompressible fluid on a $2$-dimensional surface with arbitrary topology embedded in $3$-dimensional Euclidean space by using a tailored Clebsch parametrization of the flow. In the inviscid limit, we identify conserved surface energy and enstrophy, 
		and obtain the corresponding noncanonical Hamiltonian structure. 
		We then discuss the formulation of the diffusion operator on 
		the surface by examining two alternatives. 
		In the first case, we follow the standard approach for the Navier-Stokes equations on a Riemannian manifold and 
		calculate the diffusion operator by requiring that flows corresponding to Killing fields of the Riemannian metric are not subject to dissipation. For an embedded surface, this leads to a diffusion operator including derivatives of the stream function across the surface. 
		In the second case, using an analogy with the Poisson equation for the Newtonian gravitational potential in general relativity, we construct a diffusion operator taking into account the Ricci scalar curvature of the surface. 
		The resulting vorticity equation is $2$-dimensional, and the corresponding diffusive equilibria 
		minimize dissipation under the constraint of curvature energy.
		%
		\end{abstract}

\section{Introduction}

Incompressible $2$-dimensional fluid flow can be regarded as a subsystem of the $3$-dimensional Navier Stokes equations \cite{Doering} 
in which fluid velocity is restricted to a horizontal plane.   
This setting is often encountered in geophysical fluid dynamics and plasma physics, which share governing equations with comparable mathematical structure due to the similarity between Coriolis force  and Lorentz force.  
For example, atmospheric and oceanic flows on rotating planetary surfaces are described through reduced models such as
the quasi-geostrophic equation \cite{Charney,Charney3,Horton} or the $\beta$-plane model \cite{Rhines}, 
while electrostatic turbulence in plasmas can be modeled through the Hasegawa-Mima equation \cite{HM,HM2} 
or the Hasegawa-Wakatani system \cite{HM4,Wakatani,HM3}. 

In the inviscid limit, the Navier-Stokes equations reduce to the Euler equations.  
The inviscid invariants of the Euler system are energy and helicity in $3$ dimensions, energy and enstrophy in $2$ dimensions. These invariants are associated with a noncanonical Hamiltonian structure \cite{Morrison98,Morrison82} where energy corresponds to the Hamiltonian of the system, while helicity and enstrophy arise as Casimir invariants of the respective Poisson algebras. The Hasegawa-Mima and related equations, which include the $2$-dimensional Euler system as a special case, admit an analogous Hamiltonian formulation \cite{Weinstein,Tassi}.   
Conservation of enstrophy, which originates from the vanishing of the vortex stretching term in the relevant budget equation, characterizes the behavior of $2$-dimensional turbulence.  
In contrast with the $3$-dimensional case, energy cascades toward small wavenumbers, 
while enstrophy is transferred and eventually dissipated at small spatial scales where viscous effects become dominant \cite{Kraichnan,Kraichnan2,Batchelor}. Such inverse energy cascade \cite{Rivera,Xiao}, which is observed in geophysical fluids \cite{Charney71} and turbulent plasmas \cite{HM5,Numata} as well, forms the theoretical basis of zonal flows and self-organization in $2$-dimensional turbulence \cite{Diamond}.    

The distinction between $3$-dimensional and $2$-dimensional turbulence described above hinges upon the flat geometry of Euclidean space in $3$ and $2$ dimensions. However, 
dimensionality does not depend on the specific geometry (Riemannian metric) of the manifold 
where the fluid flows. Therefore, it seems natural to ask what happens to the scenario above 
if one considers the flow of a fluid over a general $2$-diemensional surface $\Sigma$ embedded in $\mathbb{R}^3$.   
This ultimately translates into the problem of deriving a vorticity equation
for the stream function on such a surface, 
and to establish whether enstrophy is preserved. 
It is well known that special cases, such as flow on a $2$-sphere, admit a
vorticity equation for the stream function with a corresponding invariant enstrophy (see e.g. \cite{Williams,Dritschel}). Furthermore, by using a Clebsch representation \cite{Yoshida09} of the velocity field, Yoshida and Morrison \cite{Yoshida17} showed that $3$-dimensional Euler flows of the so called epi-2D form $\bol{v}=\nabla\phi+p\nabla q$, where $\phi,p,q$ are Clebsch parameters, admit an enstrophy type invariant defined over a $3$-dimensional domain originating from the vanishing of the helicity associated with such $\bol{v}$.  

Therefore, our first aim in this paper is to derive the vorticity equation for the stream function of an incompressible fluid over an arbitrary smooth surface $\Sigma\subset\mathbb{R}^3$, to obtain the preserved enstrophy of the system, and to describe the associated noncanonical Hamiltonian structure. This is done in sections $2$, $3$, and $4$. 

A complete description of $2$-dimensional fluid flows requires an accurate modeling of
dissipation. On an embedded curved surface this task is difficult because it is not clear 
how the curvature of the surface 
affects the microscopic process driving diffusion. 
The standard appraoch for Navier-Stokes flow on $n$-dimensional Riemannian manifolds \cite{Chan,Samavaki}
consists in replacing the advective derivative with the covariant derivative (in the inviscid limit this 
ensures conservation of energy and, for $n=3$, helicity \cite{Peralta16}), and in 
deriving the diffusion operator by requiring that Killing fields of the Riemannian metric (tangent vectors along which the Riemannian distance between points remains unchanged) 
are not subject to dissipation \cite{Chan,Samavaki,Ebin}. This latter condition is motivated by analogy with 
the Laplacian operator encountered in Euclidean space, and it produces a vorticity equation consistent with conservation of angular momentum on the $2$-sphere \cite{Williams,Obuse,Silberman}.  
However, it should be noted that a spherical surface shares the same rotational Killing field as the Euclidean metric of $\mathbb{R}^3$, and therefore it cannot be used to test the validity of the construction for a general surface.  Indeed, the diffusion operator on the $2$-sphere can be obtained by substituting the stream function $\Psi=R^2\xi\lr{\theta,\phi,t}$, with $\lr{R,\theta,\phi}$ spherical coordinates, into the standard Laplacian in $\mathbb{R}^3$ (see section 2 on this point).
Furthermore, while the approach based on Killing fields may be natural on an isolated (not embedded) manifold, 
it also leads to the standard Laplacian as candidate diffusion operator on any surface embedded in $\mathbb{R}^3$   
because microscopic interactions should not be modified by the embedding.  
Nonetheless, the expression of the standard Laplacian 
on a general surface does not always result in a $2$-dimensional equation 
because derivatives of the stream function in the direction across the surface appear.     
This conflicts with the physical expectation that the diffusion process should be independent of gradients across the surface since the fluid cannot escape from it.  

Hence, our second purpose in this study is to derive a diffusion operator 
on a general surface using a different approach not relying on the Killing fields of the Riemannian metric. The idea is that in the limit of vanishing Ricci scalar curvature (the case of a flat surface), the outcome of the diffusion process should be a constant vorticity on the surface, and that the diffusion operator should be consistent with the spherical case discussed above. We will see that this construction leads to an analogy with the Poisson equation for the gravitational potential in general relativity.  

The present paper is organized as follows. 
In section 2 we derive the vorticity equation for the stream function of an incompressible fluid on an arbitrary $2$-dimensional smooth surface $\Sigma\subset\mathbb{R}^3$.
In section 3 the derived equation is written through the curvilinear coordinates spanning $\Sigma$. In section 4 we obtain the noncanonical Hamiltonian structure associated with the inviscid limit.
In section 5 we discuss the relationship between the derived vorticity equation and incompressible Euler flow on a $2$-dimensional Riemannian manifold.
Section 6 is dedicated to the construction of the diffusion operator.
In section 7, we discuss examples of diffusive equilibria of the derived equation over a $2$-sphere and a $2$-torus with axial symmetry and circular cross section.
Conclusions are drawn in section 8. 

Finally, a remark on the notation used in the paper.
Coordinates will be specified either with an upper numerical index $\lr{x^1,x^2,x^3,...}$,
or through different symbols, e.g. $\lr{\mu,\nu,\zeta,...}$. 
Cartesian coordinates in $3$-dimensional Euclidean space will be expressed as $\bol{x}=\lr{x,y,z}$. 
Given a tensor, we will denote its components either through numerical indices
or directly by appending the coordinate symbols, for example $T^{12}$ or $T^{\mu\nu}$.   
Similarly, tangent basis vectors $\p\bol{x}/\p x^i$ will be expressed either as $\lr{\p_1,\p_2,\p_3,...}$ or $\lr{\p_{\mu},\p_{\nu},\p_{\zeta},...}$, 
and summation on repeated indices will be used.






\section{Vorticity equation on a surface}
Consider the incompressible Navier-Stokes equations 
\begin{subequations}
\begin{align}
\frac{\p\bol{u}}{\p t}=&-\bol{u}\cdot\nabla\bol{u}-\frac{1}{\rho}\nabla P+\frac{\sigma}{\rho}\Delta\bol{u},~~~~\nabla\cdot\bol{u}=0,\label{NS0}\\
\frac{\p\rho}{\p t}=&-\nabla\cdot\lr{\rho\bol{u}}.\label{CEq}
\end{align}\label{NS}
\end{subequations}
Here, $\bol{u}\lr{\bol{x},t}$ denotes the fluid velocity, $\rho\lr{\bol{x},t}$ the fluid density, $P\lr{\bol{x},t}$ the pressure, and $\sigma>0$ the dynamic viscosity. 
We are interested in the properties of equation \eqref{NS} when the fluid velocity $\bol{u}$
is tangential to a 2-dimensional surface $\Sigma\subset\mathbb{R}^3$ corresponding to a level set of a smooth function $\zeta\lr{\bol{x}}$. 
The unit outward normal $\bol{n}$ to the surface $\Sigma$ can be written in the form
\begin{equation}
    \bol{n}=\frac{\nabla\zeta}{\abs{\nabla\zeta}}.\label{nor}
\end{equation}
It is convenient to introduce a curvilinear coordinate system $\lr{x^1,x^2,x^3}=\lr{\mu,\nu,\zeta}$ with tangent basis $\lr{\p_{\mu},\p_{\nu},\p_{\zeta}}$ 
and cotangent basis $\lr{\nabla\mu,\nabla\nu,\nabla\zeta}$. Then, the requirement  $\bol{u}\cdot\bol{n}=0$ that the fluid velocity lies on the surface $\Sigma$ implies that
\begin{equation}
    \bol{u}=u^{\mu}\p_{\mu}+u^{\nu}\p_{\nu},\label{u}
\end{equation}
where $u^{\mu}$ and $u^{\nu}$ are the contravariant components of the field $\bol{u}$ in the $\p_{\mu}$ and $\p_{\nu}$ directions. 
In the following we shall assume that the pressure is barotropic, $P=P\lr{\rho}$, and that the density is constant on $\Sigma$, $\rho=\rho\lr{\zeta}$. 
Under these hypothesis, the continuity equation \eqref{CEq} is identically satisfied.
Let $J=\nabla\mu\cdot\nabla\nu\cp\nabla\zeta$ denote the Jacobian determinant of the coordinate transformation $\lr{x,y,z}\leftrightarrow\lr{\mu,\nu,\zeta}$. 
Substituting equation \eqref{u} into the incompressibility condition $\nabla\cdot\bol{u}=0$ of \eqref{NS}, one obtains
\begin{equation}
    \frac{\p }{\p\mu}\lr{\frac{u^{\mu}}{J}}+\frac{\p }{\p\nu}\lr{\frac{u^{\nu}}{J}}=0.
\end{equation}
This implies
\begin{equation}
    u^{\mu}=J\frac{\p\Psi}{\p\nu},~~~~u^{\nu}=-J\frac{\p\Psi}{\p\mu},~~~~\Psi=-\int\frac{u^{\nu}}{J}d\mu,\label{psi}
\end{equation}
where $\Psi$ is the stream function. 
Combining this expression with equation \eqref{u}, the fluid velocity 
can now be expressed as
\begin{equation}
    \bol{u}=J\lr{\frac{\p\Psi}{\p\nu}\p_{\mu}-\frac{\p\Psi}{\p\mu}\p_{\nu}}=\nabla\Psi\cp\nabla\zeta.\label{u2}
\end{equation}
Next, denote with $\bol{\omega}=\nabla\cp\bol{u}$ the vorticity of the fluid, 
and take the curl of the first equation in system \eqref{NS0},
\begin{equation}
    \frac{\p\bol{\omega}}{\p t}=\nabla\cp\lr{\bol{u}\cp\bol{\omega}}+\frac{\sigma}{\rho}\Delta\bol{\omega}+\frac{\sigma}{\rho^2}\frac{\p\rho}{\p\zeta}\nabla\zeta\cp\lr{\nabla\cp\bol{\omega}}.\label{omega}
\end{equation}
Since the fluid velocity $\bol{u}$ lies on surfaces of constant $\zeta$, we wish to extract 
the evolution of the contravariant component of $\bol{\omega}$ in the $\zeta$-direction,
\begin{equation}
    \omega^{\zeta}=\bol{\omega}\cdot\nabla\zeta=-\nabla\cdot\left[\nabla\zeta\cp\lr{\nabla\Psi\cp\nabla\zeta}\right],\label{omzeta}
\end{equation}
from equation \eqref{omega}. Taking the scalar product of equation \eqref{omega} with $\nabla\zeta$, using standard vector identities, and recalling that $\bol{u}\cdot\nabla\zeta=0$, we have
\begin{equation}
    \frac{\p\omega^{\zeta}}{\p t}
    =-\bol{u}\cdot\nabla\omega^{\zeta}+\frac{\sigma}{\rho}\Delta\bol{\omega}\cdot\nabla\zeta.\label{omega2}
\end{equation}
It is convenient to introduce the bracket notation
\begin{equation}
    \left[f,g\right]=\frac{\p f}{\p\mu}\frac{\p g}{\p\nu}-\frac{\p f}{\p\nu}\frac{\p g}{\p\mu}.\label{bra}
\end{equation}
Using \eqref{bra}, equation \eqref{omega2} becomes
\begin{equation}
\frac{\p\omega^{\zeta}}{\p t}=J\left[\Psi,\omega^{\zeta}\right]+\frac{\sigma}{\rho}\Delta\bol{\omega}\cdot\nabla\zeta.\label{omega3}    
\end{equation}
Or, defining the differential operator $\Delta_{\perp\zeta}$ according to
\begin{equation}
    \Delta_{\perp\zeta}\Psi=\nabla\cdot\left[\nabla\zeta\cp\lr{\nabla\Psi\cp\nabla\zeta}\right]
\end{equation}
in terms of the stream function $\Psi$ one has
\begin{equation}
    \frac{\p}{\p t}\Delta_{\perp\zeta}\Psi=J\left[\Psi,\Delta_{\perp\zeta}\Psi\right]-\frac{\sigma}{\rho}\left[\Delta\nabla\cp\lr{\nabla\Psi\cp\nabla\zeta}\right]\cdot\nabla\zeta.\label{omega4}
\end{equation}
The operator $\Delta_{\perp\zeta}$ is called a normal Laplacian \cite{SatoNormal}, in the sense that it gives the divergence of the normal component of $\nabla\Psi$ with respect to $\nabla\zeta$, weighted by a factor $\abs{\nabla\zeta}^2$.

Let us verify that for an inviscid flow with $\sigma=0$, both energy and enstrophy are preserved on each surface $\Sigma$. First, we define the energy through the functional 
\begin{equation}
    H_{\Sigma}=\frac{1}{2}\int_{\Sigma}\bol{u}^2\frac{d\mu d\nu}{J}=\frac{1}{2}\int_{\Sigma}\abs{\nabla\Psi\cp\nabla\zeta}^2\frac{d\mu d\nu}{J}.\label{E1}
\end{equation}
The rate of change in $H_{\Sigma}$ is therefore
\begin{equation}
\begin{split}
    \frac{dH_{\Sigma}}{dt}=&\int_{\Sigma}\left\{\nabla\cdot\left[\Psi\nabla\zeta\cp\lr{\nabla\frac{\p\Psi}{\p t}\cp\nabla\zeta}\right]-\Psi\frac{\p}{\p t}\Delta_{\perp\zeta}\Psi\right\}\frac{d\mu d\nu}{J}\\
    =&\int_{\Sigma}\left\{
    \frac{\p}{\p\mu}\left[\frac{\Psi}{J}\nabla\zeta\cp\lr{\nabla\frac{\p\Psi}{\p t}\cp\nabla\zeta}\cdot\nabla\mu\right]+
    \frac{\p}{\p\nu}\left[\frac{\Psi}{J}\nabla\zeta\cp\lr{\nabla\frac{\p\Psi}{\p t}\cp\nabla\zeta}\cdot\nabla\nu\right]
    -\frac{1}{2}\left[\Psi^2,\Delta_{\perp\zeta}\Psi\right]\right\}d\mu d\nu\\
    =&
    \int_{\Sigma}
    \frac{\p}{\p\mu}\left[\frac{\Psi}{J}\nabla\zeta\cp\lr{\nabla\frac{\p\Psi}{\p t}\cp\nabla\zeta}\cdot\nabla\mu-\frac{\Psi^2}{2}\frac{\p\Delta_{\perp\zeta}\Psi}{\p\nu}\right]d\mu d\nu\\&+\int_{\Sigma}
    \frac{\p}{\p\nu}\left[\frac{\Psi}{J}\nabla\zeta\cp\lr{\nabla\frac{\p\Psi}{\p t}\cp\nabla\zeta}\cdot\nabla\nu+\frac{\Psi^2}{2}\frac{\p\Delta_{\perp\zeta}\Psi}{\p\mu}\right]d\mu d\nu.
    \end{split}
\end{equation}
If the surface $\Sigma$ is closed (e.g. a 2-sphere or a 2-torus) this integral
identically vanishes because the integrand only contains partial derivatives with respect to $\mu$ and $\nu$ (this corresponds to periodic boundary conditions). Otherwise, it becomes zero upon appropriate choice of boundary conditions, such as $\Psi=0$ on $\p\Sigma$, with $\p\Sigma$ the boundary of $\Sigma$. 
Next, we consider the following definition for the enstrophy,
\begin{equation}
    W_{\Sigma}=\frac{1}{2}\int_{\Sigma}\lr{\omega^{\zeta}}^2\frac{d\mu d\nu}{J}=\frac{1}{2}\int_{\Sigma}\lr{\Delta_{\perp\zeta}\Psi}^2\frac{d\mu d\nu}{J}.\label{enst}
\end{equation}
We have
\begin{equation}
    \frac{dW_{\Sigma}}{dt}=\frac{1}{2}\int_{\Sigma}\left[\Psi,\lr{\Delta_{\perp\zeta}\Psi}^2\right]d\mu d\nu=\frac{1}{2}\int_{\Sigma}\left\{\frac{\p}{\p\mu}\left[\Psi\frac{\p\lr{\Delta_{\perp\zeta}\Psi}^2}{\p\nu}\right]-\frac{\p}{\p\nu}\left[\Psi\frac{\p\lr{\Delta_{\perp\zeta}\Psi}^2}{\p\mu}\right]\right\}d\mu d\nu.
\end{equation}
This integral vanishes under the same circumstances described above for the energy $H_{\Sigma}$. It should also be noted that, more generally, 
$W_{\Sigma}$ remains a constant even if the integrand is replaced with $f\lr{\omega^{\zeta}}/J$, with $f$ some function of $\omega^{\zeta}$. 

To conclude this section, let us verify that equation \eqref{omega4}
correctly reduces to the flow over a 2-sphere when $\zeta=R$ is the sphere radius. 
Let $\lr{R,\theta,\phi}$ denote a spherical coordinate system with $\theta\in\left[0,\pi\right]$ the polar
angle and $\phi\in\left[0,2\pi\right)$ the azimuthal angle. 
Then, one can verify that the contravariant radial component of the vorticity is given by 
\begin{equation}
  \omega^{R}=-\Delta_{\perp R}\Psi=-\Delta\Psi+\frac{1}{R^2}\frac{\p}{\p R}\lr{R^2\frac{\p\Psi}{\p R}}=-\frac{1}{R^2\sin\theta}\left[\frac{\p}{\p\theta}\lr{\sin\theta\frac{\p\Psi}{\p\theta}}+\frac{1}{\sin\theta}\frac{\p^2\Psi}{\p\phi^2}\right]=-\frac{1}{R^2}\Delta_S \Psi,  
\end{equation}
where $\Delta_S$ denotes the Laplace-Beltrami operator on the 2-sphere. Next, observe that the diffusion term in equation \eqref{omega4} can be expressed as
\begin{equation}
\begin{split}
    \left[\Delta\nabla\cp\lr{\nabla\Psi\cp\nabla R}\right]&\cdot\nabla R=\nabla\cp\nabla\cp\nabla\cp\lr{\sin\theta\frac{\p\Psi}{\p\theta}\nabla\phi-\frac{1}{\sin\theta}\frac{\p\Psi}{\p\phi}\nabla\theta}\cdot\nabla R\\
    =&\nabla\cp\nabla\cp\lr{\frac{\Delta_{S}\Psi}{R^2}\nabla R-\frac{\p^2\Psi}{\p R\p\phi}\nabla\phi-\frac{\p^2\Psi}{\p R\p\theta}\nabla\theta}\cdot\nabla R\\
    =&\nabla\cdot\left\{\left[\nabla\cp\lr{\frac{\Delta_S\Psi}{R^2}
    \nabla R
    -\frac{\p^2\Psi}{\p R\p\theta}\nabla\theta-\frac{\p^2\Psi}{\p R\p\phi}\nabla\phi
    }\right]\cp\nabla R\right\}\\
    =&\nabla\cdot\left[-R^2\frac{\p}{\p R}\lr{\frac{\Delta_S\Psi}{R^2}}\nabla\lr{\frac{1}{R}}
    -\nabla\frac{\Delta_S\Psi}{R^2}-\frac{\p^3\Psi}{\p  R^2\p\phi}\nabla\phi+\sin\theta\frac{\p^3\Psi}{\p R^2\p\theta}\nabla\lr{\int\frac{d\theta}{\sin\theta}}
    \right]\\
    =&\frac{1}{R^2}\frac{\p}{\p R}\lr{R^2\frac{\p}{\p R}\frac{\Delta_S\Psi}{R^2}}-\Delta\frac{\Delta_S\Psi}{R^2}-\frac{1}{R^2}\lr{\frac{1}{\sin^2\theta}\frac{\p^4\Psi}{\p R^2\p\phi^2}+\frac{1}{\sin\theta}\frac{\p}{\p\theta}\sin\theta\frac{\p^3\Psi}{\p R^2\p\theta}}\\
    =&-\frac{1}{R^4}\Delta_{S}^2\Psi-\frac{1}{R^2}\frac{\p^2\Delta_{S}\Psi}{\p R^2}.
\end{split}\label{LBS}
\end{equation}
Hence, setting $\mu=\theta$ and $\nu=\phi$, equation \eqref{omega4} reduces to
\begin{equation}
    \frac{\p}{\p t}\Delta_S\Psi=\frac{1}{R^2\sin\theta}\left[\Psi,\Delta_{S}\Psi\right]+\frac{\sigma}{\rho}\left(\frac{\Delta_S^2\Psi}{R^2}+\frac{\p^2\Delta_S\Psi}{\p R^2}\right). 
\end{equation}
Notice that in this context it is customary to demand that $\Psi=R^2\xi\lr{\theta,\phi,t}$ and $\rho=\rho_0/R^2$, with $\rho_0$ a positive constant, 
to eliminate the radial dependence of the equation. 
In such case one obtains a $2$-dimensional equation 
\begin{equation}
    \frac{\p}{\p t}\Delta_S\xi=\frac{1}{\sin\theta}\left[\xi,\Delta_{S}\xi\right]+\frac{\sigma}{\rho_0}\lr{\Delta_{S}^2\xi+2\Delta_S\xi},\label{SphereD}
\end{equation}
which is the usual form of the vorticity equation over a 2-sphere.

\section{Vorticity equation in curvilinear coordinates}
The purpose of the present section is to 
express the evolution of the stream function $\Psi$ in a form that is more appropriate for applications, such as numerical simulations.
Indeed, equation \eqref{omega4} 
reflects the embedding of the surface $\Sigma$ in Euclidean space $\mathbb{R}^3$, while it is desirable to write the vorticity equation 
only through $\Psi$ and its derivatives with respect to $\mu$ and $\nu$. 
The disentanglement of equation \eqref{omega4} from the 3-dimensional representation can be pursued by introducing the covariant and contravariant metric tensors
\begin{equation}
    g_{ij}=\p_i\cdot\p_j,~~~~g^{ij}=\nabla x^i\cdot\nabla x^j,~~~~i,j=1,2,3,\label{g}
\end{equation}
In this notation $\lr{x^1,x^2,x^3}=\lr{\mu,\nu,\zeta}$ and $\lr{\p_1,\p_2,\p_3}=\lr{\p_{\mu},\p_{\nu},\p_{\zeta}}$. Notice that $g_{ij}$ is the inverse matrix of $g^{ij}$, that is $g_{ij}g^{jk}=\delta_i^{~k}$.
Furthermore, $J=1/\sqrt{\abs{g}}$, with $\abs{g}$ the determinant of the covariant metric tensor $g_{ij}$. 
Using \eqref{g}, the vorticity \eqref{omzeta} can be expressed as
\begin{equation}
    \omega^{\zeta}=-\Delta_{\perp\zeta}\Psi=-J\left\{\frac{\p}{\p\mu}\left[
    J\lr{g_{\nu\nu}\frac{\p\Psi}{\p\mu}-g_{\mu\nu}\frac{\p\Psi}{\p\nu}}
    \right]
    +\frac{\p}{\p\nu}\left[
    J\lr{g_{\mu\mu}\frac{\p\Psi}{\p\nu}-g_{\mu\nu}\frac{\p\Psi}{\p\mu}}
    \right]
    \right\}.\label{omzetapsi}
\end{equation}
For an orthogonal coordinate system the above expression can be further simplified since the metric tensor is diagonal. In such case, 
\begin{equation}
    \omega^{\zeta}=-\Delta_{\perp\zeta}\Psi=-J\left[\frac{\p}{\p\mu}\lr{Jg_{\nu\nu}\frac{\p\Psi}{\p\mu}}
    +\frac{\p}{\p\nu}\lr{Jg_{\mu\mu}\frac{\p\Psi}{\p\nu}}
    \right].\label{omzetaperp}
\end{equation}
With the help of equation \eqref{omzetapsi} one can then write the advective term $J\left[\Psi,\omega^{\zeta}\right]$ in \eqref{omega3} through $\Psi$ and its derivatives with respect to $\mu$ and $\nu$. 
The handling of the diffusion term is more involved. First, observe that the
vorticity corresponding to the flow \eqref{u} is
\begin{equation}
    \begin{split}
    \bol{\omega}=&
    J\left\{
    \frac{\p}{\p\nu}
    \left[J\lr{g_{\mu\zeta}\frac{\p\Psi}{\p\nu}-g_{\nu\zeta}\frac{\p\Psi}{\p\mu}}\right]-\frac{\p}{\p\zeta}
    \left[J
    \lr{g_{\mu\nu}\frac{\p\Psi}{\p\nu}-g_{\nu\nu}\frac{\p\Psi}{\p\mu}}
\right]
    \right\}\p_{\mu}\\
    &J\left\{
\frac{\p}{\p\zeta}
\left[J\lr{g_{\mu\mu}\frac{\p\Psi}{\p\nu}
-g_{\mu\nu}\frac{\p\Psi}{\p\mu}}
\right]
-\frac{\p}{\p\mu}\left[
J\lr{
g_{\mu\zeta}\frac{\p\Psi}{\p\nu}
-g_{\nu\zeta}\frac{\p\Psi}{\p\mu}
}
\right]
    \right\}\p_{\nu}\\
    &J\left\{
    \frac{\p}{\p\mu}\left[
    J\lr{g_{\mu\nu}\frac{\p\Psi}{\p\nu}-g_{\nu\nu}\frac{\p\Psi}{\p\mu}}
        \right]
        -\frac{\p}{\p\nu}\left[
        J\lr{g_{\mu\mu}\frac{\p\Psi}{\p\nu}
        -g_{\mu\nu}\frac{\p\Psi}{\p\mu}}
        \right]
    \right\}\p_{\zeta}.\label{vorticity}
\end{split}
\end{equation}
Next, it is convenient to write the diffusion term in \eqref{omega3} as follows:
\begin{equation}
\Delta\bol{\omega}\cdot\nabla\zeta=-\nabla\cdot\left(\bol{\xi}\cp\nabla\zeta\right)=J\lr{\frac{\p\xi_{\mu}}{\p\nu}-\frac{\p\xi_{\nu}}{\p\mu}},~~~~\bol{\xi}=\nabla\cp\bol{\omega}.\label{diff1}
\end{equation}
The right-hand side of the first equation can be evaluated by taking the curl of \eqref{vorticity}. 
For a general curvilinear coordinate system the resulting expression is lengthy and it is therefore omitted here. For practical purposes it is however useful to consider the special case of orthogonal coordinates. 
If $\lr{\mu,\nu,\zeta}$ defines an orthogonal coordinate system, the vorticity \eqref{vorticity} takes the form
\begin{equation}
\begin{split}
\bol{\omega}=&
    J\frac{\p}{\p\zeta}\lr{Jg_{\nu\nu}\frac{\p\Psi}{\p\mu}}\p_{\mu}+
    J
\frac{\p}{\p\zeta}
\left(Jg_{\mu\mu}\frac{\p\Psi}{\p\nu}
\right)
    \p_{\nu}
    -J\left[
    \frac{\p}{\p\mu}\left(
    Jg_{\nu\nu}\frac{\p\Psi}{\p\mu}
        \right)
        +\frac{\p}{\p\nu}\left(
        Jg_{\mu\mu}\frac{\p\Psi}{\p\nu}
        \right)
    \right]\p_{\zeta},\label{vorticity2}
\end{split}
\end{equation}
so that the diffusion term \eqref{diff1} reads as
\begin{equation}
\begin{split}
\Delta\bol{\omega}\cdot\nabla\zeta=&-J\frac{\p}{\p\mu}\left\{
\frac{J}{g^{\nu\nu}}\frac{\p}{\p\zeta}
\left[
\frac{J}{g^{\mu\mu}}
\frac{\p}{\p\zeta}\lr{Jg_{\nu\nu}\frac{\p\Psi}{\p\mu}}
\right]
+\frac{J}{g^{\nu\nu}}\frac{\p}{\p\mu}
\left[
\frac{J}{g^{\zeta\zeta}}\frac{\p}{\p\mu}\lr{Jg_{\nu\nu}\frac{\p\Psi}{\p\mu}}+\frac{J}{g^{\zeta\zeta}}\frac{\p}{\p\nu}\lr{Jg_{\mu\mu}\frac{\p\Psi}{\p\nu}}
\right]
\right\}
\\
&-J\frac{\p}{\p\nu}
\left\{
\frac{J}{g^{\mu\mu}}\frac{\p}{\p\zeta}
\left[
\frac{J}{g^{\nu\nu}}\frac{\p}{\p\zeta}\lr{Jg_{\mu\mu}\frac{\p\Psi}{\p\nu}}\right]
+\frac{J}{g^{\mu\mu}}\frac{\p}{\p\nu}
\left[
\frac{J}{g^{\zeta\zeta}}\frac{\p}{\p\mu}\lr{Jg_{\nu\nu}\frac{\p\Psi}{\p\mu}}
+\frac{J}{g^{\zeta\zeta}}\frac{\p}{\p\nu}\lr{Jg_{\mu\mu}\frac{\p\Psi}{\p\nu}}
\right]
\right\}
\\=&
-J\frac{\p}{\p\mu}\left\{
\frac{J}{g^{\nu\nu}}\frac{\p}{\p\zeta}
\left[
\frac{J}{g^{\mu\mu}}
\frac{\p}{\p\zeta}\lr{Jg_{\nu\nu}\frac{\p\Psi}{\p\mu}}
\right]\right\}
-J\frac{\p}{\p\nu}
\left\{
\frac{J}{g^{\mu\mu}}\frac{\p}{\p\zeta}
\left[
\frac{J}{g^{\nu\nu}}\frac{\p}{\p\zeta}\lr{Jg_{\mu\mu}\frac{\p\Psi}{\p\nu}}\right]\right\}\\
&+J\left\{
\frac{\p}{\p\mu}\left[
Jg_{\nu\nu}\frac{\p}{\p\mu}\lr{\frac{\omega^{\zeta}}{g^{\zeta\zeta}}}
\right]
+\frac{\p}{\p\nu}\left[Jg_{\mu\mu}\frac{\p}{\p\nu}\lr{\frac{\omega^{\zeta}}{g^{\zeta\zeta}}}\right]
\right\}\\
=&-J\frac{\p}{\p\mu}\left\{
\frac{J}{g^{\nu\nu}}\frac{\p}{\p\zeta}
\left[
\frac{J}{g^{\mu\mu}}
\frac{\p}{\p\zeta}\lr{Jg_{\nu\nu}\frac{\p\Psi}{\p\mu}}
\right]\right\}
-J\frac{\p}{\p\nu}
\left\{
\frac{J}{g^{\mu\mu}}\frac{\p}{\p\zeta}
\left[
\frac{J}{g^{\nu\nu}}\frac{\p}{\p\zeta}\lr{Jg_{\mu\mu}\frac{\p\Psi}{\p\nu}}\right]\right\}\\&+\Delta_{\perp\zeta}\lr{\frac{\omega^{\zeta}}{g^{\zeta\zeta}}}\\
=&-\mc{D}\Psi+\Delta_{\perp\zeta}\lr{\frac{\omega^{\zeta}}{g^{\zeta\zeta}}},\label{omzetaperp2}
\end{split}
\end{equation}
where we used \eqref{omzetaperp} and introduced the fourth-order differential operator
\begin{equation}
\mc{D}\Psi=J\frac{\p}{\p\mu}\left\{
\frac{J}{g^{\nu\nu}}\frac{\p}{\p\zeta}
\left[
\frac{J}{g^{\mu\mu}}
\frac{\p}{\p\zeta}\lr{Jg_{\nu\nu}\frac{\p\Psi}{\p\mu}}
\right]\right\}
+J\frac{\p}{\p\nu}
\left\{
\frac{J}{g^{\mu\mu}}\frac{\p}{\p\zeta}
\left[
\frac{J}{g^{\nu\nu}}\frac{\p}{\p\zeta}\lr{Jg_{\mu\mu}\frac{\p\Psi}{\p\nu}}\right]\right\}.\label{dPsi}
\end{equation}
Now the vorticity equation \eqref{omega3} takes the form
\begin{equation}
\frac{\p\omega^{\zeta}}{\p t}=J\left[\Psi,\omega^{\zeta}\right]+\frac{\sigma}{\rho}\left[-\mc{D}\Psi+\Delta_{\perp\zeta}\lr{\frac{\omega^{\zeta}}{g^{\zeta\zeta}}}\right].\label{vor2d}
\end{equation}
From this expression one sees that 
the diffusion term contains derivatives of $\Psi$ with respect to $\zeta$, implying that in the presence of viscosity ($\sigma\neq 0$) the vorticity equation is not 2-dimensional for a general coordinate system $\lr{\mu,\nu,\zeta}$. 
A sufficient condition for the diffusion term \eqref{omzetaperp2} to be 2-dimensional is that $\p g_{\nu\nu}/\p\zeta=\p g_{\mu\mu}/\p\zeta=\p g_{\zeta\zeta}/\p\zeta=0$ so that one may set $\p\Psi/\p\zeta=0$ in the vorticity equation. These conditions on the metric coefficients are satisfied when $\p_{\zeta}=\bol{a}+\bol{b}\cp\bol{x}$, $\bol{a},\bol{b}\in\mathbb{R}^3$, is the generator of continuous Euclidean isometries (rotations and translations) in $\mathbb{R}^3$, because they are the only continuous transformation preserving the Euclidean distance between points. 
Another possibility to achieve a 2-dimensional equation not containing derivatives of $\Psi$ with respect to $\zeta$ is to make an appropriate ansatz for the functional dependence of $\Psi$ and $\rho$ on $\zeta$ as in the case of spherical coordinates discussed in the previous section. Nevertheless, 
the 3-dimensional obstruction represented by the term $-\mc{D}\Psi$ suggests that
the 3-diemsnional Laplacian operator $\Delta$ may be physically inappropriate to describe viscous dissipation on an embedded 2-dimensional surface. 
These aspects will be further discussed in section 6, where the formulation of a diffusion operator consistent with surface curvature is studied.  


\section{Noncanonical Hamiltonian structure}
In this section we derive the Hamiltonian structure associated with the vorticity equation \eqref{omega4} in the inviscid limit $\sigma=0$.
In the following we shall restrict our attention to the case in which the surface $\Sigma$ is closed, 
and the system is periodic in the variables $\mu$ and $\nu$. 
The Hamiltonian of the system is given by the energy functional \eqref{E1}. 
Using \eqref{omzeta}, the Hamiltonian \eqref{E1} can be equivalently expressed as follows
\begin{equation}
    \begin{split}
    H_{\Sigma}
    =&\frac{1}{2}\int_{\Sigma}\left\{\nabla\cdot\left[\Psi\nabla\zeta\cp\lr{\nabla\Psi\cp\nabla\zeta}\right]+\Psi\omega^{\zeta}\right\}\frac{d\mu d\nu}{J}\\
    =&\frac{1}{2}\int_{\Sigma}\left\{\frac{\p}{\p\mu}\left[
    \frac{\Psi\nabla\zeta\cp\lr{\nabla\Psi\cp\nabla\zeta}\cdot\nabla\mu}{J}
    \right]+\frac{\p}{\p\nu}\left[
    \frac{\Psi\nabla\zeta\cp\lr{\nabla\Psi\cp\nabla\zeta}\cdot\nabla\nu}{J}
    \right]\right\}d\mu d\nu\\&+\frac{1}{2}\int_{\Sigma}\Psi\omega^{\zeta}\frac{d\mu d\nu}{J}
    \end{split}
\end{equation}
Since $\Sigma$ is a closed surface, 
it follows that
\begin{equation}
    H_{\Sigma}=\frac{1}{2}\int_{\Sigma}\Psi\omega^{\zeta}\frac{d\mu d\nu}{J}.
\end{equation}
To calculate functional derivatives, we must define the inner product between functions on $\Sigma$. In the following, we consider the standard $L^2\lr{\Sigma}$ inner product defined with respect to the measure $d\mu d\nu$ on $\Sigma$. 
Next, observe that the variation of $H_{\Sigma}$ 
with respect to the field $\omega^{\zeta}$ is
\begin{equation}
\begin{split}
    \delta_{\omega^{\zeta}}H_{\Sigma}=&
    \frac{1}{2}\int_{\Sigma}\lr{-\delta\Psi\Delta_{\perp\zeta}\Psi+\Psi\delta\omega^{\zeta}}\frac{d\mu d\nu}{J}\\
    =&\frac{1}{2}\int_{\Sigma}\left\{
    \nabla\cdot\left[
    \Psi\nabla\zeta\cp\lr{\nabla\delta\Psi\cp\nabla\zeta}-\delta\Psi\nabla\zeta\cp\lr{\nabla\Psi\cp\nabla\zeta}
    \right]
    +2\Psi\delta\omega^{\zeta}
    \right\}\frac{d\mu d\nu}{J}.
\end{split}
\end{equation}
Again, using periodic boundary conditions, one obtains
\begin{equation}
    \frac{\delta H_{\Sigma}}{\delta\omega^{\zeta}}=\frac{\Psi}{J}.
\end{equation}
Given two functionals $F,G$ of the field $\omega^{\zeta}$, we now consider the Poisson bracket,
\begin{equation}
    \left\{F,G\right\}=\int_{\Sigma}\omega^{\zeta}\left[J\frac{\delta F}{\delta\omega^{\zeta}},J\frac{\delta G}{\delta\omega^{\zeta}}\right]d\mu d\nu=-\int_{\Sigma}J\frac{\delta F}{\delta\omega^{\zeta}}\left[\omega^{\zeta},J\frac{\delta G}{\delta\omega^{\zeta}}\right]d\mu d\nu.\label{PB}
\end{equation}
which must be validated by verifying the relevant axioms of antisymmetry and Jacobi identity (see e.g. \cite{Morrison98,Olver,Morrison82}). The antisymmetry
\begin{equation}
    \left\{F,G\right\}=-\left\{G,F\right\}
\end{equation} of \eqref{PB} is immediate, since it follows from the antisymmetry of the bracket $\left[,\right]$.
Similarly, given three functionals $F,G,H$, the Jacobi identity 
\begin{equation}
    \left\{F,\left\{G,H\right\}\right\}+\left\{G,\left\{H,F\right\}\right\}+\left\{H,\left\{F,G\right\}\right\}=0,\label{JI}
\end{equation}
for \eqref{PB} follows from that associated with the anti-self-adjoint operator
\begin{equation}
\mf{J}=-J\left[\omega^{\zeta},J\circ\right]. 
\end{equation} 
The anti-self-adjointness is given by the property
\begin{equation}
    \int_{\Sigma}f\mf{J}g\,d\mu d\nu=-\int_{\Sigma}g\mf{J}f\,d\mu d\nu
\end{equation}
for any pair of functions $f,g$. 
Notice also that in terms of $\mf{J}$, the Poisson bracket \eqref{PB} reads as
\begin{equation}
    \left\{F,G\right\}=\int_{\Sigma}\frac{\delta F}{\delta\omega^{\zeta}}\mf{J}\frac{\delta G}{\delta\omega^{\zeta}}\,d\mu d\nu.
\end{equation}
To verify the Jacobi identity, observe that
\begin{equation}
    \frac{\delta\left\{G,H\right\}}{\delta\omega^{\zeta}}=\left[J\frac{\delta G}{\delta\omega^{\zeta}},J\frac{\delta H}{\delta\omega^{\zeta}}\right]
    -J\frac{\delta^2 G }{\delta\lr{\omega^{\zeta}}^2}\left[\omega^{\zeta},J\frac{\delta H}{\delta\omega^{\zeta}}\right]
    +J\frac{\delta^2 H }{\delta\lr{\omega^{\zeta}}^2}\left[\omega^{\zeta},J\frac{\delta G}{\delta\omega^{\zeta}}\right].
\end{equation}
This implies,
\begin{equation}
    \begin{split}
    \left\{F,\left\{G,H\right\}\right\}=&
    \int_{\Sigma}\omega^{\zeta}\left[J\frac{\delta F}{\delta\omega^{\zeta}},
    J\left[J\frac{\delta G}{\delta\omega^{\zeta}},J\frac{\delta H}{\delta\omega^{\zeta}}\right]
    \right]\,d\mu d\nu\\
    &-\int_{\Sigma}\left\{\omega^{\zeta}\left[J\frac{\delta F}{\delta\omega^{\zeta}},J^2\frac{\delta^2 G}{\delta\lr{\omega^{\zeta}}^2}\left[
    \omega^{\zeta},J\frac{\delta H}{\delta\omega^{\zeta}}
    \right]-J^2\frac{\delta^2 H }{\delta\lr{\omega^{\zeta}}^2}\left[\omega^{\zeta},J\frac{\delta G}{\delta\omega^{\zeta}}\right]\right]\label{JI2}\right\}\,d\mu d\nu.
\end{split}
\end{equation}
Upon substitution of this expression into the Jacobi identity \eqref{JI}, contributions originating from the first term on the right-hand side of equation \eqref{JI2} 
vanish due to the Jacobi identity satisfied by the bracket $\left[,\right]$. Indeed, for any triplet of functions $f,g,h$, one has
\begin{equation}
    \left[f,\left[g,h\right]\right]+\left[g,\left[h,f\right]\right]+\left[h,\left[f,g\right]\right]=0.
\end{equation}
Therefore, denoting with $\circlearrowright$ summation on even permutations, 
\begin{equation}
\int_{\Sigma}\omega^{\zeta}\left[J\frac{\delta F}{\delta\omega^{\zeta}},J\left[J\frac{\delta G}{\delta\omega^{\zeta}},J\frac{\delta H}{\delta\omega^{\zeta}}\right]\right]\,d\mu d\nu+\circlearrowright
=\int_{\Sigma}\omega^{\zeta}\left[J\frac{\delta 
F}{\delta\omega^{\zeta}},J\right]
\left[J\frac{\delta G}{\delta\omega^{\zeta}},J\frac{\delta H}{\delta\omega^{\zeta}}\right]
\, d\mu d\nu+\circlearrowright
=0.
\end{equation}
Terms involving second order variations in \eqref{JI2} cancel due to the anti-self-adjointness of $\mf{J}$. 
For example, the first term in the second integral on the right-hand side of equation \eqref{JI2} 
cancels with the following term appearing in $\left\{H,\left\{F,G\right\}\right\}$, 
\begin{equation}
\begin{split}
    \int_{\Sigma}\omega^{\zeta}\left[
    J\frac{\delta H}{\delta\omega^{\zeta}},J^2\frac{\delta^2 G}{\delta\lr{\omega^{\zeta}}^2}\left[
    \omega^{\zeta},J\frac{\delta F}{\delta\omega^{\zeta}}
    \right]
    \right]\,d\mu d\nu=&-\int_{\Sigma}J\frac{\delta H}{\delta\omega^{\zeta}}\left[\omega^{\zeta},J^2\frac{\delta^2G}{\delta\lr{\omega^{\zeta}}^2}\left[\omega^{\zeta},J\frac{\delta F}{\delta\omega^{\zeta}}\right]\right]d\mu d\nu\\
    =&\int_{\Sigma}J^2\frac{\delta^2 G}{\delta\lr{\omega^{\zeta}}^2}\left[\omega^{\zeta},J\frac{\delta F}{\delta\omega^{\zeta}}\right]\left[\omega^{\zeta},J\frac{\delta H}{\delta \omega^{\zeta}}\right]\,d\mu d\nu\\
    =&-\int_{\Sigma}J\frac{\delta  F}{\delta\omega}\left[\omega^{\zeta},J^2\frac{\delta^2 G}{\delta\lr{\omega^{\zeta}}^2}\left[\omega^{\zeta},J\frac{\delta H}{\delta \omega^{\zeta}}\right]\right]\,d\mu d\nu\\
    =&\int_{\Sigma}\omega^{\zeta}\left[J\frac{\delta F}{\delta\omega^{\zeta}},J^2\frac{\delta^2 G}{\delta\lr{\omega^{\zeta}}^2}\left[
    \omega^{\zeta},J\frac{\delta H}{\delta\omega^{\zeta}}
    \right]\right]\,d\mu d\nu.
\end{split}
\end{equation}
where we used the anti-self-adjointness of $\mf{J}$.
Hence, the Jacobi identity is satsifed, and \eqref{PB} is a Poisson bracket. 
Noting that
\begin{equation}
    \left\{\omega^{\zeta},H_{\Sigma}\right\}=-\int_{\Sigma}J\delta\lr{\mu-\mu'}\delta\lr{\nu-\nu'}\left[\omega^{\zeta},\Psi\right]d\mu d\nu=J\left[\Psi,\omega^{\zeta}\right],
\end{equation}
the vorticity equation \eqref{omega3} in the inviscid limit $\sigma=0$ takes the noncanonical Hamiltonian form 
\begin{equation}
    \frac{\p\omega^{\zeta}}{\p t}=\left\{\omega^{\zeta},H_{\Sigma}\right\}.
\end{equation}
Finally, observe that the Poisson bracket \eqref{PB} is endowed with the Casimir invariant
\begin{equation}
    C_{\Sigma}=\int_{\Sigma}f\lr{\omega^{\zeta}}\frac{d\mu d\nu}{J},
\end{equation}
where $f\lr{\omega^{\zeta}}$ is some function of $\omega^{\zeta}$. 
Indeed, 
\begin{equation}
    \left\{C_{\Sigma},H_{\Sigma}\right\}=0~~~~\forall H_{\Sigma}.
\end{equation}
The enstrophy \eqref{enst} corresponds to the choice $f\lr{\omega^{\zeta}}=\lr{\omega^{\zeta}}^2$. 

\section{Relation with fluid flow on a Riemannian manifold}

Consider the ideal flow of an incompressible fluid over an n-dimensional Riemannian manifold $M$ with coordinates $\bol{x}=\lr{x^1,...,x^n}$,
\begin{equation}
    \frac{\p\bol{u}}{\p t}+\nabla_{\bol{u}}\bol{u}=-{\rm grad}\, P,~~~~{\rm div}\,\bol{u}=0.\label{IERM}
\end{equation}
Here, the density is assumed constant, and the gradient and divergence operators are defined with respect to the Riemannian metric,
\begin{equation}
    {\rm grad}\,P=\frac{\p P}{\p x^i}g^{ij}\p_j,~~~~{\rm div}\,\bol{u}=\frac{1}{\sqrt{\abs{g}}}\frac{\p}{\p x^i}\lr{\sqrt{\abs{g}}u^i},
\end{equation}
with $\abs{g}$ the determinant of the covariant metric tensor $g_{ij}$. Furthermore,  
$\nabla_{\bol{u}}$ denotes the covariant derivative,
\begin{equation}
    \nabla_{\bol{u}}\bol{v}=\lr{u^i\frac{\p v^k}{\p x^i}+u^iv^j\Gamma_{ij}^k}\p_k,
\end{equation}
where the Christoffel symbols of the second kind $\Gamma_{ij}^k$  are related to the metric coefficients  according to  
\begin{equation}
    \Gamma_{ij}^k=\frac{1}{2}g^{km}\lr{\frac{\p g_{m i}}{\p x^j}+\frac{\p g_{mj}}{\p x^i}-\frac{\p g_{ij}}{\p x^m}},~~~~i,j,k=1,...,n.
\end{equation}
Christoffel symbols of the first kind are defined as $\Gamma_{ijk}=g_{im}\Gamma^{m}_{jk}$. 
It should be noted that equation \eqref{IERM} preserves the energy 
\begin{equation}
H_M=\frac{1}{2}\int_{M} u^iu^jg_{ij}\sqrt{\abs{g}}\,dx^1...dx^n\label{E2}
\end{equation} 
for all $n$, and, when $n=3$, the helicity (see  \cite{Peralta16} for the definition of helicity). 
Let us verify that $H_M$ is indeed constant. 
Recalling that ${\rm div}\,\bol{u}=0$, we have
\begin{equation}
\begin{split}
\frac{dH_M}{dt}=&-\int_{M}u^ig_{ij}\lr{
u^k\frac{\p u^j}{\p x^k}+u^ku^\ell\Gamma_{k\ell}^j
+P_kg^{kj}
}\sqrt{\abs{g}}\,dx^1...dx^n\\
=&-\int_{M}\left[\frac{\p}{\p x^i}\lr{P\sqrt{\abs{g}}u^i}+u^ig_{ij}u^k\frac{\p u^j}{\p x^k}\sqrt{\abs{g}}+\frac{1}{2}u^iu^ku^{\ell}\lr{
\frac{\p g_{ik}}{\p x^{\ell}}+
\frac{\p g_{i\ell}}{\p x^{k}}-
\frac{\p g_{k\ell}}{\p x^{i}}
}\sqrt{\abs{g}}\right]\,dx^1...dx^n\\
=&-\int_{M}\left[\frac{\p}{\p x^i}\lr{P\sqrt{\abs{g}}u^i}+\lr{u^kg_{kj}u^i\frac{\p u^j}{\p x^i}+\frac{1}{2}u^iu^ku^{\ell}
\frac{\p g_{ik}}{\p x^{\ell}}
}\sqrt{\abs{g}}\right]\,dx^1...dx^n\\
=&-\int_{M}\left\{\frac{\p}{\p x^i}\left[u^i\lr{P+u_ju^j}\sqrt{\abs{g}}\right]
+\left[-u^iu^j\frac{\p}{\p x^i}\lr{u^kg_{kj}}+\frac{1}{2}u^iu^ku^{\ell}
\frac{\p g_{ik}}{\p x^{\ell}}
\right]\sqrt{\abs{g}}\right\}\,dx^1...dx^n\\
=&-\int_{M}\left\{\frac{\p}{\p x^i}\left[u^i\lr{P+\frac{1}{2}u_ju^j
}\sqrt{\abs{g}}
\right]
\right\}\,dx^1...dx^n\\
&-\int_{M}\left\{
\left[
\frac{1}{2}u^iu^kg_{kj}\frac{\p u^j}{\p x^i}
-\frac{1}{2}u^iu^j\frac{\p}{\p x^i}\lr{u^kg_{kj}}
+\frac{1}{2}u^iu^ju^k\frac{\p g_{ik}}{\p x^{\ell}}
\right]
\right\}\,dx^1...dx^n\\
=&-\int_{M}\left\{\frac{\p}{\p x^i}\left[u^i\lr{P+\frac{1}{2}u_ju^j
}\sqrt{\abs{g}}
\right]
\right\}\,dx^1...dx^n\\
=&-\int_{M}{\rm div}\left[\bol{u}\lr{P+\frac{1}{2}\abs{u}^2}\right]\sqrt{\abs{g}}\,dx^1...dx^n.
\end{split}
\end{equation}
Here, we introduced the notation $\abs{u}^2=u^ju_j$. 
From the last passage, one sees that the rate of change in energy can be written as a boundary integral which vanishes under appropriate boundary conditions. 

We now wish to determine how the derived vorticity equation \eqref{omega3} with $\sigma=0$ relates to equation \eqref{IERM}
when the Riemannian manifold is a 2-dimensional surface $\Sigma$, i.e. $n=2$.
First, note that for a 2-dimensional flow $\bol{u}=u^{\mu}\p_{\mu}+u^{\nu}\p_{\nu}$ over a surface $\Sigma$ embedded in 3-dimensional Euclidean space $\mathbb{R}^3$, 
the first equation in system \eqref{IERM} is equivalent to
\begin{equation}
    \frac{\p\bol{u}}{\p t}=-\frac{\nabla\zeta\cp\left[\lr{\nabla P+\bol{u}\cdot\nabla\bol{u}}\cp\nabla\zeta\right]}{\abs{\nabla\zeta}^2}.\label{omega5}
\end{equation}
Indeed, the covariant derivative and the gradient operators in \eqref{IERM} effectively correspond to the projections of the usual $\mathbb{R}^3$ operators over the surface $\Sigma$ through the projector $-\abs{\nabla\zeta}^{-2}\nabla\zeta\cp\lr{\nabla\zeta\cp}$.
Taking the curl of both sides in \eqref{omega5} gives 
\begin{equation}
    \frac{\p\bol{\omega}}{\p t}=-\nabla\cp
    \left\{\frac{\nabla\zeta\cp\left[
    \lr{\nabla P+\bol{u}\cdot\nabla\bol{u}}\cp\nabla\zeta
    \right]}{\abs{\nabla\zeta}^2}\right\}.
\end{equation}
On the other hand, the second equation in system \eqref{IERM} implies that 
\begin{equation}
\bol{u}=\frac{1}{\sqrt{\abs{g}}}\lr{\frac{\p\Psi}{\p\nu}\p_{\mu}-\frac{\p\Psi}{\p\mu}\p_{\nu}},\label{usp}
\end{equation}
where $\Psi$ is the stream function. 
The contravariant $\zeta$-component $\omega^{\zeta}=\bol{\omega}\cdot\nabla\zeta$ of the vorticity therefore obeys
\begin{equation}
    \frac{\p\omega^{\zeta}}{\p t}=
    -\bol{u}\cdot\nabla\omega^{\zeta}
    -\nabla\cdot\left[\nabla\lr{P+\frac{\bol{u}^2}{2}}\cp\nabla\zeta\right]=\frac{1}{\sqrt{\abs{g}}}\left[\Psi,\omega^{\zeta}\right].\label{omega_sp}
\end{equation}
Hence, the inviscid case $\sigma=0$ of \eqref{omega3} differs from the equation above due to the discrepancy between the $3$-dimensional Jacobian $J=\nabla\mu\cdot\nabla\nu\cp\nabla\zeta$ and the surface area  $\sqrt{\abs{g}}=\abs{\p_{\mu}\cp\p_{\nu}}=\abs{\nabla\zeta}/J$. This discrepancy reflects the 
dependence of the notion of divergence on the volume element: in equation \eqref{omega3} the divergence
is inherited from $\mathbb{R}^3$, while in \eqref{omega_sp} is defined with respect to the metric of the $2$-dimensional surface. Notice also that this discrepancy is absent, for example, on a spherical surface since $\abs{\nabla\zeta}=\abs{\nabla R}=1$. 

\section{Viscous dissipation on a surface with arbitrary topology}
In section 3 we have seen that the diffusion term in the vorticity equation \eqref{vor2d}
\begin{equation}
-\mc{D}\Psi+\Delta_{\perp\zeta}\lr{\frac{\omega^{\zeta}}{g^{\zeta\zeta}}},
\end{equation}
arising from the Navier-Stokes equations in 3-dimensional Euclidean space contains derivatives of $\Psi$ with respect to $\zeta$ due to the term $\mc{D}\Psi$. 
Hence, the vorticity equation \eqref{vor2d} 
cannot be generally regarded as a 2-dimensional equation on the surface $\Sigma$. 
Furthermore, the 3-dimensional Laplacian operator $\Delta$ does not take into account the Riemannian curvature of the underlying manifold (recall that Euclidean space $\mathbb{R}^3$ is flat and the corresponding Riemannian curvature is zero). 
However, if the fluid flow is restricted to a 2-dimensional surface $\Sigma\subset\mathbb{R}^3$, the corresponding diffusion operator  
should be consistent with the topology of the surface, in the sense that derivatives of $\Psi$ across the surface do not appear and the effect of curvature on the diffusion process is taken into account. 
The aim of this section is to attempt the derivation of  
such consistent diffusion operator for a viscous flow on an arbitrary 2-dimensional surface $\Sigma$ embedded in $\mathbb{R}^3$. 
We will examine two alternatives. The first case is the conventional approach of \cite{Ebin,Chan} for Riemannian manifolds in which the diffusion operator is derived  
by generalizing a mathematical property of the Laplacian operator in $\mathbb{R}^3$, namely that Killing fields of the Riemannian metric are not subject to diffusion (they belong to the kernel of the diffusion operator). In the second alternative, the diffusion operator is constructed by 
translating the expected physical outcome of the diffusion process for the vorticity $\omega^{\zeta}$ 
into a target functional whose
stationary points assign the 
diffusive equilbria 
of the system. 
In particular, the diffusion operator is obtained by demanding that 
for a purely diffusive process the vorticity $\omega^{\zeta}$ becomes constant on any flat $2$-dimensional surface at equilibrium.

We begin by considering the first setting. 
Observe that the vorticity equation in $\mathbb{R}^3$ (equation \eqref{omega}) admits stationary solutions of the form
\begin{equation}
\bol{u}=\bol{\chi},~~~~\rho=\rho_0,~~~~\bol{\chi}=\bol{a}+\bol{b}\cp\bol{x},~~~~\bol{a},\bol{b}\in\mathbb{R}^3,~~~~\rho_0\in\mathbb{R}.
\end{equation}
Furthermore, the action of the Laplacian operator $\Delta$ on the velocity field above returns zero, 
\begin{equation}
\Delta\bol{\chi}=-\nabla\cp\nabla\cp\lr{\bol{b}\cp\bol{x}}=\bol{0}, 
\end{equation}
implying that $\bol{\chi}$ does not undergo diffusion. 
On the other hand, recall that $\bol{\chi}$ is the Killing field associated with the Euclidean metric of $\mathbb{R}^3$. Indeed, it preserves
the Euclidean distance between points,
\begin{equation}
\mf{L}_{\bol{\chi}}g_{ij}dx^i dx^j=\lr{g_{jk}\frac{\p\chi^k}{\p x^i}+g_{ki}\frac{\p\chi^k}{\p x^j}+\frac{\p g_{ij}}{\p x^k}\chi^k}dx^idx^j=0.\label{Kill1}
\end{equation}
Here, $\mf{L}$ denotes the Lie derivative, $\lr{x^1,x^2,x^3}$ some curvilinear coordinate system, and $g_{ij}$ the corresponding covariant metric tensor.  
In Cartesian coordinates $\lr{x^1,x^2,x^3}=\lr{x,y,z}$, equation \eqref{Kill1} takes the form
\begin{equation}
\frac{\p\chi^j}{\p x^i}+\frac{\p \chi^i}{\p x^j}=0,~~~~i,j=1,2,3.\label{Kill2}
\end{equation}
Evidently, the Killing field $\bol{u}=\bol{\chi}$ corresponds to a minimum of the energy dissipation functional
\begin{equation}
\mc{U}=\frac{1}{4}\sum_{i,j=1}^3\int_{\mathbb{R}^3}\lr{\frac{\p u^j}{\p x^i}+\frac{\p u^i}{\p x^j}}^2\,dx^1dx^2dx^3.
\end{equation}
Furthermore, recalling that $\nabla\cdot\bol{u}=0$, 
\begin{equation}
\frac{\delta\mc{U}}{\delta u^i}=-\Delta u^i,~~~~i=1,2,3.\label{U}
\end{equation}
This equation gives a relationship between the Killing field $\bol{\chi}$ of the Euclidean metric and the diffusion operator $\Delta$ appearing in the Navier-Stokes system. 
The next objective is to derive a
diffusion operator for fluid flow on a 2-dimensional surface such that the 
Killing fields of the corresponding metric tensor correspond to stationary solutions of the vorticity equation, in analogy with the 3-dimensional case discussed above. 
To this end, notice that on a Riemannian manifold $M$ of diemnsion $n$  
the covariant derivative of a Killing field $\bol{\chi}$ with respect to itself can be written through the gradient operator,
\begin{equation}
\begin{split}
\nabla_{\bol{\chi}}\bol{\chi}=&\left[\chi^ig_{k\ell}\frac{\p\chi^k}{\p x^i}+\frac{1}{2}\chi^i\chi^j\lr{\frac{\p g_{i\ell}}{\p x^j}+\frac{\p g_{\ell i}}{\p x^j}-\frac{\p g_{ij}}{\p x^{\ell}}}\right]\nabla x^{\ell}\\
=&\left[\chi^ig_{k\ell}\frac{\p\chi^k}{\p x^i}+\chi^i\chi^j\frac{\p g_{i\ell}}{\p x^j}
+g_{ij}\frac{\p\chi^i}{\p x^\ell}\chi^j
\right]\nabla x^{\ell}-{\rm grad}\lr{\frac{1}{2}\chi^i\chi^jg_{ij}}\\
=&-\frac{1}{2}{\rm grad}\abs{\bol{\chi}}^2,
\end{split}
\end{equation}
where we used the Killing equation \eqref{Kill1} in the last passage, and set $\abs{\bol{\chi}}^2=g_{ij}\chi^i\chi^j$. 
Therefore, if the diffusion operator is constructed in a such a way that Killing fields are not subject to diffusion, 
they will automatically correspond to stationary solutions of the corresponding Navier-Stokes system upon setting $\bol{u}=\bol{\chi}$ and $P=\abs{\bol{\chi}}^2/2$. 
Next, observe that in $n$ spatial dimensions the Killing equation \eqref{Kill1} can be equivalently written as
\begin{equation}
K_{ij}=\lr{\nabla_i\bol{u}}_j+\lr{\nabla_j\bol{u}}_i=0,~~~~i,j=1,...,n.
\end{equation}
In this notation $\nabla_i=\nabla_{\p_i}$ denotes the covariant derivative with respect to the ith tangent vector $\p_i$, i.e.
\begin{equation}
\nabla_i\bol{u}=\lr{\frac{\p u^k}{\p x^i}+u^{j}\Gamma^{k}_{ij}}\p_k.
\end{equation}
It follows that the generalization of the energy dissipation function \eqref{U} to an $n$-dimensional Riemannian manifold $M$ is
\begin{equation}
\mc{U}=\frac{1}{4}\int_{M}K_{ij}g^{ik}g^{j\ell}K_{k\ell}\sqrt{\abs{g}}\,dx^1...dx^n.\label{U2}
\end{equation}
Taking variations of \eqref{U2} with respect to $\bol{u}$ gives
\begin{equation}
\begin{split}
\delta_{\bol{u}}\mc{U}=&\frac{1}{2}\int_{M}K_{ij}g^{ik}g^{j\ell}\delta K_{k\ell}\sqrt{\abs{g}}\,dx^1...dx^n\\
=&\int_{M}\left[
g^{ik}\lr{\frac{\p u^{\ell}}{\p x^i}+u^{p}\Gamma_{ip}^{\ell}}+g^{j\ell}\lr{\frac{\p u^k}{\p x^j}+u^p\Gamma^k_{jp}}
\right]\lr{
\frac{\p\delta u^m}{\p x^k}g_{m\ell}+\Gamma_{\ell k m}\delta u^m
}\sqrt{\abs{g}}\,dx^1...dx^n\\
=&-\int_{M}\delta u^m
\left\{
\frac{\p}{\p x^k}\left[
\sqrt{\abs{g}}\lr{g^{ik}\frac{\p u_m}{\p x^i}-u^{\ell}g^{ik}\lr{\frac{\p g_{m\ell}}{\p x^i}-
\Gamma_{m i\ell}-\Gamma_{im\ell}
}+\frac{\p u^k}{\p x^m}}
\right]
\right\}
\,dx^1...dx^n\\
&+\int_{M}\delta u^m\left[
\frac{\p u^{\ell}}{\p x^i}g^{ik}\lr{\Gamma_{\ell k m}+\Gamma_{k\ell m}}+u^{j}g^{ik}\Gamma_{ij}^{\ell}\lr{\Gamma_{\ell k m}+\Gamma_{k\ell m}}
\right]\sqrt{\abs{g}}\,dx^1...dx^n\\
=&\int_{M}\delta u^m\left\{
-\Delta_{M}u_m+\frac{1}{\sqrt{\abs{g}}}\frac{\p}{\p x^k}\left[\sqrt{\abs{g}}u^{\ell}g^{ik}\lr{\frac{\p g_{m\ell}}{\p x^i}-\frac{\p g_{im}}{\p x^{\ell}}}
\right]+u^k\frac{\p\Gamma_{m \ell}^{\ell}}{\p x^k}
\right\}\sqrt{\abs{g}}\,dx^1...dx^n\\
&+\int_{M}\delta u^m\lr{\frac{\p u^\ell}{\p x^i}g^{ik}\frac{\p g_{k\ell}}{\p x^m}+u^{\ell}g^{ik}\Gamma_{i\ell}^{j}\frac{\p g_{jk}}{\p x^{m}}}\sqrt{\abs{g}}\,dx^1...dx^n\\
=&\int_{M}\delta u^m\left\{
-\Delta_{M}u_m+u^{\ell}\left[
-2\Gamma_{m\ell}^k\Gamma_{kj}^{j}-2\frac{\p\Gamma_{m\ell}^{k}}{\p x^k}+\frac{1}{\sqrt{\abs{g}}}\frac{\p}{\p x^k}\lr{\sqrt{\abs{g}}g^{ik}\frac{\p g_{i\ell}}{\p x^m}}+\frac{\p\Gamma_{mk}^{k}}{\p x^\ell}
\right]
\right\}\sqrt{\abs{g}}\, dx^1...dx^n
\\&+\int_{M}\delta u^m\left[2\frac{\p u^{\ell}}{\p x^k}g^{ik}\Gamma_{\ell i m}+u^{\ell}\Gamma_{i\ell}^j\lr{
2\Gamma^{i}_{jm}-g^{ik}\lr{\frac{\p g_{km}}{\p x^j}-\frac{\p g_{jm}}{\p x^k}}
}\right]\sqrt{\abs{g}}\,dx^1...dx^n\\
=&\int_{M}\delta u^m\lr{-\Delta_{M}u_m-2u^{\ell}R_{\ell m}+2\frac{\p u^{\ell}}{\p x^k}g^{ik}\Gamma_{\ell im}}\sqrt{\abs{g}}\,dx^1...dx^n\\
&+
\int_{M}\delta u^m\left\{u^{\ell}\left[
\frac{1}{\sqrt{\abs{g}}}\frac{\p}{\p x^k}\lr{\sqrt{\abs{g}}g^{ik}\frac{\p g_{i\ell}}{\p x^m}}
-\frac{\p\Gamma_{mk}^{k}}{\p x^{\ell}}+\Gamma_{i\ell}^jg^{ik}\lr{\frac{\p g_{jm}}{\p x^k}-\frac{\p g_{km}}{\p x^j}}
\right]
\right\}\sqrt{\abs{g}}\, dx^1...dx^n.
\end{split}\label{mcU}
\end{equation}
Here, we used ${\rm div}\bol{u}=0$, $\p\sqrt{\abs{g}}/\p x^k=\sqrt{\abs{g}}\Gamma_{kp}^p$, and the vanishing of variations on boundaries. Furthermore, $\Delta_{M}$ denotes the Laplace-Beltrami operator
\begin{equation}
\Delta_{M}=\frac{1}{\sqrt{\abs{g}}}\frac{\p}{\p x^i}\lr{\sqrt{\abs{g}}g^{ik}\frac{\p}{\p x^k}},
\end{equation}
while $R_{\ell m}$ is the Ricci curvature tensor,
\begin{equation}
R_{\ell m}=\frac{\p\Gamma_{\ell m}^{\sigma}}{\p x^{\sigma}}-\frac{\p\Gamma_{m\sigma}^{\sigma}}{\p x^{\ell}}+\Gamma^{i}_{i\sigma}\Gamma_{\ell m}^{\sigma}-\Gamma_{\ell i}^{\sigma}\Gamma_{\sigma m}^{i}.
\end{equation}
Hence, from \eqref{mcU} we obtain
\begin{equation}
\frac{\delta\mc{U}}{\delta u^m}=-\Delta_{M}u_m-\lr{2R_{\ell m}-\chi_{\ell m}}u^\ell+2\frac{\p u^{\ell}}{\p x^k}g^{ik}\Gamma_{\ell im},\label{NSD1}
\end{equation}
where we introduced the covariant tensor
\begin{equation}
\chi_{\ell m}=\frac{1}{\sqrt{\abs{g}}}\frac{\p}{\p x^k}\lr{\sqrt{\abs{g}}g^{ik}\frac{\p g_{i\ell}}{\p x^m}}
-\frac{\p\Gamma_{mk}^{k}}{\p x^{\ell}}+\Gamma_{i\ell}^jg^{ik}\lr{\frac{\p g_{jm}}{\p x^k}-\frac{\p g_{km}}{\p x^j}}.
\end{equation}
The constant density Navier-Stokes equations on an $n$-dimensional Riemannian manifold $M$ including the diffusion operator \eqref{NSD1} therefore read as
\begin{equation}
\frac{\p\bol{u}}{\p t}+\nabla_{\bol{u}}\bol{u}=-{\rm grad} P-\sigma\frac{\delta\mc{U}}{\delta \bol{u}},~~~~{\rm div}\bol{u}=0.\label{NSD12}
\end{equation}
In this notation, $\frac{\delta\mc{U}}{\delta\bol{u}}=\frac{\delta\mc{U}}{\delta u^m}g^{m\ell}\p_{\ell}$.
Recalling that for a $2$-dimensional surface $M=\Sigma\subset\mathbb{R}^3$ the fluid velocity is given by 
\eqref{usp}, 
one can express equation \eqref{NSD12} solely in terms of $\Psi$, and construct a corresponding $2$-dimensional vorticity equation.
At this point, it is useful to verify the consistency of the diffusion operator \eqref{NSD1} with the
spherical case studied in section 2, equation \eqref{LBS}. Indeed, 
the Killing field $\bol{\xi}=\bol{b}\cp\bol{x}$ associated with rotations in $\mathbb{R}^3$ 
is also a Killing field with respect to the metric tensor on a spherical surface, 
and we therefore expect the projection of the usual diffusion operator on the sphere 
\begin{equation}
\mc{D}_{\perp R}\bol{u}=\nabla R\cp\lr{\Delta\bol{u}\cp\nabla R}, 
\end{equation}
to coincide with \eqref{NSD1} when $\Psi=R^2\xi\lr{\theta,\phi,t}$. Considering the surface $R=1$, one can verify that
\begin{equation}
\mc{D}_{\perp R}\bol{u}=\left[\Delta_{\Sigma}u_{\theta}+2u_{\theta}+u_{\theta}\lr{\frac{\cos^2\theta}{\sin^2\theta}-1}+2\frac{\cos\theta}{\sin\theta}\frac{\p u_{\theta}}{\p\theta}\right]\nabla \theta+\left[\Delta_{\Sigma}u_{\phi}+2u_{\phi}+2\frac{\cos\theta}{\sin\theta}\lr{\frac{\p u_{\theta}}{\p\phi}-\frac{\p u_{\phi}}{\p\theta}}\right]\nabla\phi.
\end{equation}
Furthermore, we have
\begin{equation}
-\lr{\chi_{\ell m}u^{\ell}+2\frac{\p u^{\ell}}{\p x^k}g^{ik}\Gamma_{\ell im}}g^{mp}\p_p=\left[u_{\theta}\lr{\frac{\cos^2\theta}{\sin^2\theta}-1}+2\frac{\cos\theta}{\sin\theta}\frac{\p  u_{\theta}}{\p\theta}\right]\nabla \theta+\left[2\frac{\cos\theta}{\sin\theta}\lr{\frac{\p u_{\theta}}{\p\phi}-\frac{\p u_{\phi}}{\p\theta}}\right]\nabla\phi.
\end{equation}
Recalling that a sphere is an Einstein manifold such that $R_{\ell m}=g_{\ell m}/R^2$, the diffusion operators $-\delta\mc{U}/\delta \bol{u}$ and $\mc{D}_{\perp R}\bol{u}$ therefore coincide. 

The construction described above relies on the assumption that
Killing fields of the Riemannian metric should not be subject to diffusion.
While the resulting diffusion operator \eqref{NSD1} represents a consistent mathematical generalization of the
standard Laplacian to arbitrary Riemannian manifolds, 
the embedding of 
a surface $\Sigma$ in $\mathbb{R}^3$ 
should not affect the original diffusion process in $\mathbb{R}^3$ because the 
nature of the physical interactions driving dissipation are expected to be independent 
of such topological constraints. This implies that, on $\Sigma$, the argument based on Killing fields leads to the Laplacian $\Delta$ of $\mathbb{R}^3$ restricted to $\Sigma$ as diffusion operator.  
However, as shown in equation \eqref{dPsi}, the standard Laplacian results in a diffusion operator
including derivatives of the stream function across the surface, a fact that 
prevents the vorticity equation from being $2$-dimensional. 
Furthermore, if the standard Laplacian is taken, 
then only Killing fields on $\Sigma$ inherited from $\mathbb{R}^3$ should 
be free of diffusion. Notice that surfaces inheriting a continuous Euclidean isometry from $\mathbb{R}^3$   
must be symmetric with respect to such transformation. Indeed, 
suppose that the metric induced on the surface $\Sigma$ inherits the Killing field $\bol{\chi}=\bol{a}+\bol{b}\cp\bol{x}$ from
$\mathbb{R}^3$. Then, $\bol{\chi}=\chi^{\mu}\p_{\mu}+\chi^{\nu}\p_{\nu}$, which implies $\mf{L}_{\bol{\chi}}d\zeta=0$.
An alternative to obtain a $2$-dimensional vorticity equation 
would be to discard the embedding of $\Sigma$ in $\mathbb{R}^3$, treating it as a self-standing
$2$-dimensional manifold where diffusion is driven by the operator \eqref{NSD1}. However, this approach leads to a discrepancy with the vorticity equation on an embedded surface even in the absence of dissipation due to the different meaning carried by the divergence operator (recall the difference between equation \eqref{omega3} and equation \eqref{omega_sp}). Hence, we are led to consider a third pathway 
for the description of diffusion on an embedded surface. 

In order to derive a new diffusion operator $\mf{D}_{\Sigma}$ we reason as follows. 
First, 
we consider the inviscid limit $\sigma=0$ of equation \eqref{omega3}. 
If the surface $\Sigma$ is flat, we expect the outcome of a pure diffusion process to be such that 
the vorticity along the unit vector $\nabla\zeta/\abs{\nabla\zeta}$ given by   
\begin{equation}
w=\bol{\omega}\cdot\frac{\nabla\zeta}{\abs{\nabla\zeta}}=\frac{\omega^{\zeta}}{\sqrt{g^{\zeta\zeta}}},
\end{equation}
is constant on $\Sigma$. Hence, the diffusion operator must minimize 
the squared component of the gradient of 
$w$ 
tangent to the surface $\Sigma$,  
\begin{equation}
\mc{W}_{\Sigma}=\frac{1}{2}\int_{\Sigma}\frac{\abs{\nabla\zeta\cp\nabla w}^2}{\abs{\nabla\zeta}^2}\,\frac{d\mu d\nu}{J}
\end{equation}
Taking variations $\delta w=\delta\omega^{\zeta}/\sqrt{g^{\zeta\zeta}}$ of $\mc{W}_{\Sigma}$ 
with respect to the measure $J^{-1}d\mu d\nu$, 
one arrives at the following diffusion operator for a flat surface,
\begin{equation}
\begin{split}
\mf{D}_{\Sigma}\omega^{\zeta}=-\frac{\delta\mc{W}_{\Sigma}}{\delta w}=&\nabla\cdot
\left\{
\frac{\nabla\zeta\cp\left[
\nabla w\cp\nabla\zeta
\right]}{\abs{\nabla\zeta}^2}
\right\}\\
=&\frac{1}{g^{\zeta\zeta}}\left[\Delta_{\perp\zeta}w-\nabla g^{\zeta\zeta}\cdot\nabla_{\perp\zeta}w\right],
\end{split}
\end{equation}
where we defined the orthogonal gradient operator
\begin{equation}
\nabla_{\perp\zeta}=-\frac{\nabla\zeta\cp\lr{\nabla\zeta\cp\nabla}}{\abs{\nabla\zeta}^2}.
\end{equation}
Observe that if $\zeta=z$, the  operator $\mf{D}_{\Sigma}$ reduces to the $2$-dimensional Laplacian, $\mf{D}_{\Sigma}=\p^2/\p x^2+\p^2/\p y^2$.  
Now consider the case in which the $2$-dimensional surface $\Sigma$ has curvature. 
Denoting with $\gamma_{ij}$ the metric induced on $\Sigma$ and with $\gamma^{ij}$ its inverse, this means that the corresponding Ricci scalar curvature $\mf{R}=\gamma^{ij}R_{ij}$ does not vanish. 
We further assume that the system admits  configurations without vorticity, $\bol{\omega}=\bol{0}$. This implies that, even if $\mf{R}\neq 0$, the vorticity diffusion equation admits the trivial solution $\omega^{\zeta}=0$. In the limit of small curvature, $\mf{R}<<1$, this requirement points to a steady state described by
\begin{equation}
\mf{D}_{\Sigma}\omega^{\zeta}=-\frac{f\mf{R}}{\sqrt{g^{\zeta\zeta}}}\omega^{\zeta},\label{GRana}
\end{equation}
where the function $f$ has to be determined by demanding the operator to be consistent with the diffusion operator on the sphere obtained in equation \eqref{SphereD} (recall that a sphere shares the same rotational Killing field with $\mathbb{R}^3$, and therefore the correct diffusion operator on a sphere embedded in $\mathbb{R}^3$ can be obtained from the standard Laplacian $\Delta$ 
when $\Psi=R^2\xi\lr{\theta,\phi,t}$ and $\zeta=R$).
One can verify that the ricci scalar curvature of the sphere is $\mf{R}=2/R^2$. Hence, setting $\zeta=R$, one obtains
\begin{equation}
\mf{D}_{\Sigma}\omega^{\zeta}+\frac{f\mf{R}}{\sqrt{g^{\zeta\zeta}}}\omega^{\zeta}=-\frac{1}{R^2}\Delta_{\Sigma}^2\xi-\frac{2f}{R^2}\Delta_{\Sigma}\xi,
\end{equation}
with $\Delta_{\Sigma}$ the Laplace-Beltrami operator on $\Sigma$. Comparing with the diffusion operator in \eqref{SphereD}, this implies $f=1$. 
Thus, we are led to consider the following vorticity equation on an arbitrary surface $\Sigma\subset\mathbb{R}^3$,
\begin{equation}
\frac{\p\omega^{\zeta}}{\p t}=J\left[\Psi,\omega^{\zeta}\right]+\frac{\sigma}{\rho}\lr{\mf{D}_{\Sigma}\omega^{\zeta}+\frac{\mf{R}}{\sqrt{g^{\zeta\zeta}}}\omega^{\zeta}}.\label{omegaX}
\end{equation}
It should be noted that this equation does not include partial derivatives of $\Psi$ or $\omega^{\zeta}$ with respect to the surface label $\zeta$, although the coefficients may depend on $\zeta$. Hence, it can be regarded as a $2$-dimensional equation on the surface $\Sigma$. Furthermore, it is worth observing that the diffusion operator in equation \eqref{omegaX} can be obtained from the energy dissipation functional 
\begin{equation}
\mf{W}_{\Sigma}=\frac{1}{2}\int_{\Sigma}\left[
\abs{\nabla_{\perp\zeta}\lr{\frac{\omega^{\zeta}}{\sqrt{g^{\zeta\zeta}}}}}^2-\mf{R}\frac{\omega^{\zeta}\omega^{\zeta}}{g^{\zeta\zeta}}
\right]\,\frac{d\mu d\nu}{J}\label{WDiss}
\end{equation}
by taking variations with respect to $w=\omega^{\zeta}/\sqrt{g^{\zeta\zeta}}$. Indeed, we have
\begin{equation}
\mf{D}_{\Sigma}\omega^{\zeta}+\frac{\mf{R}}{\sqrt{g^{\zeta\zeta}}}\omega^{\zeta}=-\sqrt{g^{\zeta\zeta}}\frac{\delta\mf{W}_{\Sigma}}{\delta\omega^{\zeta}}.
\end{equation}
Hence, the diffusion process minimizes $\mf{W}_{\Sigma}$. 
Finally, there is an analogy between the
Poisson-like equation of diffusive equilibrium for $\omega^{\zeta}$,
\begin{equation}
\mf{D}_{\Sigma}\omega^{\zeta}=-\frac{\mf{R}}{\sqrt{g^{\zeta\zeta}}}\omega^{\zeta},\label{DiffEq}
\end{equation}
and the equation satisfied by the component of the metric tensor $g_{00}$ of a $4$-dimensional spherically symmetric static universe $\mc{M}^4$ with spatial sections $V^3$ in general relativity \cite{LeviCivita,Frankel},
\begin{equation}
\Delta_{V^3}\sqrt{-g_{00}}=-R_0^0\sqrt{-g_{00}},
\end{equation}
where $\Delta_{V^3}$ denotes the Laplace-Beltrami operator on the spatial sections $V^3$ and $R_0^0=g^{0i}R_{i0}$. Recalling that in this context $\sqrt{-g_{00}}$ can be identified with the Newtonian gravitational potential $U$ and $-R_0^0\sqrt{g_{00}}$ with the density of mass $\rho$, we have the correspondence
\begin{equation}
\frac{\omega^{\zeta}}{\sqrt{g^{\zeta\zeta}}}\leftrightarrow U,~~~~\frac{\mf{R}}{\sqrt{g^{\zeta\zeta}}}\omega^{\zeta}\leftrightarrow \rho.
\end{equation}
Hence, the second term within the integral in \eqref{WDiss} can be interpreted as a (curvature) energy density $\rho U$, and diffusive equilibrium \eqref{DiffEq} can be seen as a configuration with minimum dissipation at fixed energy. 


\section{Diffusive equilibrium on curved surfaces}
\begin{figure}[h]
\hspace*{-0cm}\centering
\includegraphics[scale=0.36]{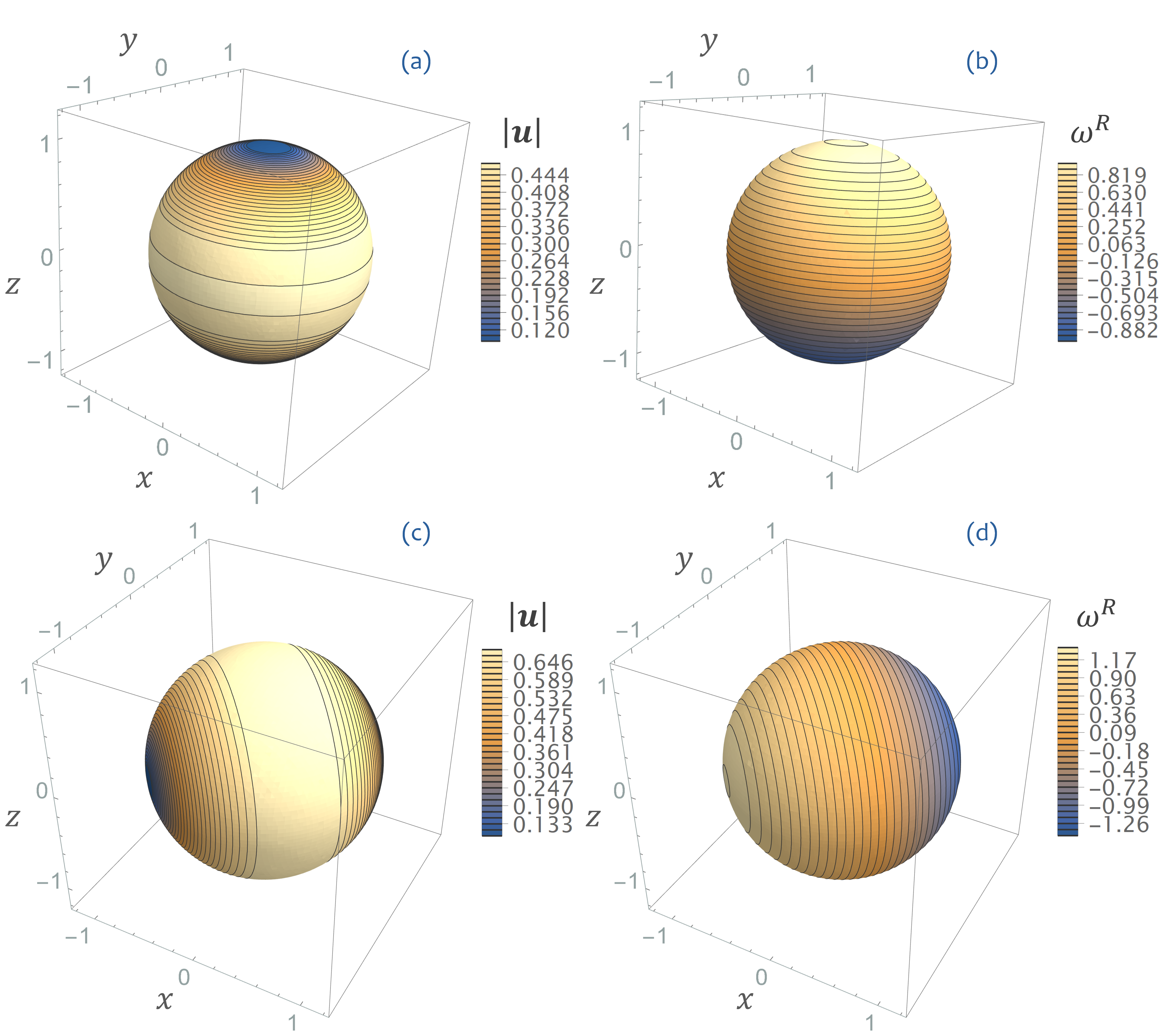}
\caption{\footnotesize (a) and (b): Contour plots of the modulus $\abs{u}$ associated with \eqref{uS1} 
and the vorticity $\omega^R$ of \eqref{omS1} for $R=B_0=1$ and $A_1=B_1=0$. Notice that in this case $\bol{u}=\p_{\phi}/2$. (c) and (d):  Contour plots of the modulus $\abs{u}$ associated with  \eqref{uS1} and the vorticity $\omega^{R}$ of \eqref{omS1} for $R=A_1=B_1=1$ and $B_0=0$.}
\label{fig1}
\end{figure}

In this section we study some examples of diffusive equilibria \eqref{DiffEq} for the vorticity $\omega^{\zeta}$.
Using the variable  $w=\omega^{\zeta}/\sqrt{g^{\zeta\zeta}}$,  
equation \eqref{DiffEq} can be explicitly written as
\begin{equation}
J\left\{\frac{\p}{\p\mu}\left[\frac{J}{g^{\zeta\zeta}}\lr{g_{\nu\nu}\frac{\p w}{\p\mu}-g_{\mu\nu}\frac{\p w}{\p\nu}}\right]
+\frac{\p}{\p\nu}\left[\frac{J}{g^{\zeta\zeta}}\lr{g_{\mu\mu}\frac{\p w}{\p\nu}-g_{\mu\nu}\frac{\p w}{\p\mu}}\right]\right\}=-\mf{R}w.\label{DE2}
\end{equation}
First, we consider the spherical case $\lr{\mu,\nu,\zeta}=\lr{\theta,\phi,R}$. Observing that $w=\omega^R$, and  that
in this context the Ricci scalar curvature is $\mf{R}=2/R^2$, equation \eqref{DE2} reduces to 
an eigenvalue problem for the Laplace-Beltrami operator on the sphere (Helmholtz equation),
\begin{equation}
\frac{1}{\sin\theta}\frac{\p}{\p\theta}\lr{\sin\theta\frac{\p \omega^R}{\p\theta}}+\frac{1}{\sin^2\theta}\frac{\p^2 \omega^R}{\p\phi^2}=-2\omega^R.\label{DES}
\end{equation}
Solutions of \eqref{DES} can be obtained by separation of variables, $\omega^R=\Theta\lr{\theta}\Phi\lr{\phi}$. The result is
\begin{equation}
\omega^R=\sum_{m=0}^{\infty}P_{\ell}^m\lr{\cos\theta}\left[A_m\sin\lr{m\phi}+B_m\cos\lr{m\phi}\right],
\end{equation}
where $P_{\ell}^m$ are associated Legendre polynomials with index $\lr{\ell,m}$, $\ell$ and $m$   positive integers, 
and $A_m,B_m\in\mathbb{R}$. 
The value of $\ell$ is determined by the curvature of the sphere 
according to  $2=\ell\lr{\ell+1}$, giving $\ell=1$. Furthermore, for $\omega^R$ to be regular at $\cos\theta=\pm 1$, 
the indices $\ell$ and $m$ must satisfy $0\leq m\leq \ell$. 
Then, the only available choices for $m$ are $m=0,1$. 
The corresponding associated Legendre polynomials are $P_{1}^0=\cos\theta$ and $P_{1}^{1}=-\sin\theta$, so that the vorticity has expression
\begin{equation}
\omega^{R}=B_0\cos\theta
-\sin\theta\left[
A_1\sin\phi+B_1\cos\phi
\right].
\label{omS1}
\end{equation}
Next, recall that 
the stream function $\Psi=R^2\xi\lr{\theta,\phi,t}$ is related to $\omega^{R}$ through equation \eqref{omzetapsi}, which in spherical coordinates gives
\begin{equation}
\frac{1}{\sin\theta}\frac{\p}{\p\theta}\lr{\sin\theta\frac{\p\xi}{\p\theta}}
+\frac{1}{\sin^2\theta}\frac{\p^2\xi}{\p \phi^2}=-\omega^{R}.\label{xiLap}
\end{equation}
Comparing equations \eqref{xiLap} and \eqref{DES}, one sees that $\xi=\omega^{R}/2$ solves \eqref{xiLap}.  
Recalling that $\bol{u}=\nabla\Psi\cp\nabla R$, one thus obtains the flow field
\begin{equation}
\bol{u}=
\frac{1}{2}\lr{B_1\sin\phi-A_1\cos\phi}\p_{\theta}+\frac{1}{2}\left[B_0+\frac{\cos\theta}{\sin\theta}\lr{A_1\sin\phi+B_1\cos\phi}\right]\p_{\phi}.\label{uS1}
\end{equation}
It should be noted that \eqref{uS1} is also a
steady solution of the full vorticity equation \eqref{omegaX} because 
$\Psi$ and $\omega^{R}$ 
commute, $\left[\Psi,\omega^R\right]=0$. 
This is expected since 
\eqref{uS1} can be identified with the  
rotational Killing field of the Euclidean metric of $\mathbb{R}^3$ and the metric induced on the sphere. Indeed, the Killing field associated with rotations is  
\begin{equation}
\bol{b}\cp\bol{x}=\lr{b_y\cos\phi-b_x\sin\phi}\p_{\theta}+\left[b_z-\frac{\cos\theta}{\sin\theta}\lr{b_x\cos\phi+b_y\sin\phi}\right]\p_{\phi},
\end{equation}
which corresponds to 
the solution \eqref{uS1} 
upon setting $b_x=-A_1/2$, $b_y=-A_2/2$, and $b_z=2B_0$.
Notice also that \eqref{uS1} holds even if the Coriolis force $-2\bol{\Omega}\cp\bol{u}$ and the centrifugal force $-\bol{\Omega}\cp\lr{\bol{\Omega}\cp\nabla R}$, with $\bol{\Omega}$ the angular velocity of the rotating frame, are included on the right-hand side of the Navier-Stokes system provided that $\bol{\Omega}=\Omega^R\nabla R$ and $\bol{u}\cdot\nabla\Omega^R=0$ with $\Omega^{R}=\bol{\Omega}\cdot\nabla R$, and in particular if $\Omega^R$ is constant. 
Indeed, the corresponding contribution to the vorticity equation is $-2\nabla R\cdot\nabla\cp\lr{\bol{\Omega}\cp\bol{u}}=-2\bol{u}\cdot\nabla \Omega^{R}$. Figure \ref{fig1} shows a plot of the vorticity $\omega^{R}$ of equation \eqref{omS1} and the modulus  $\abs{u}$ 
corresponding to the flow \eqref{uS1} for different values of the constants $B_0$, $A_1$, and $B_1$.  

\begin{figure}[h]
\hspace*{-0cm}\centering
\includegraphics[scale=0.32]{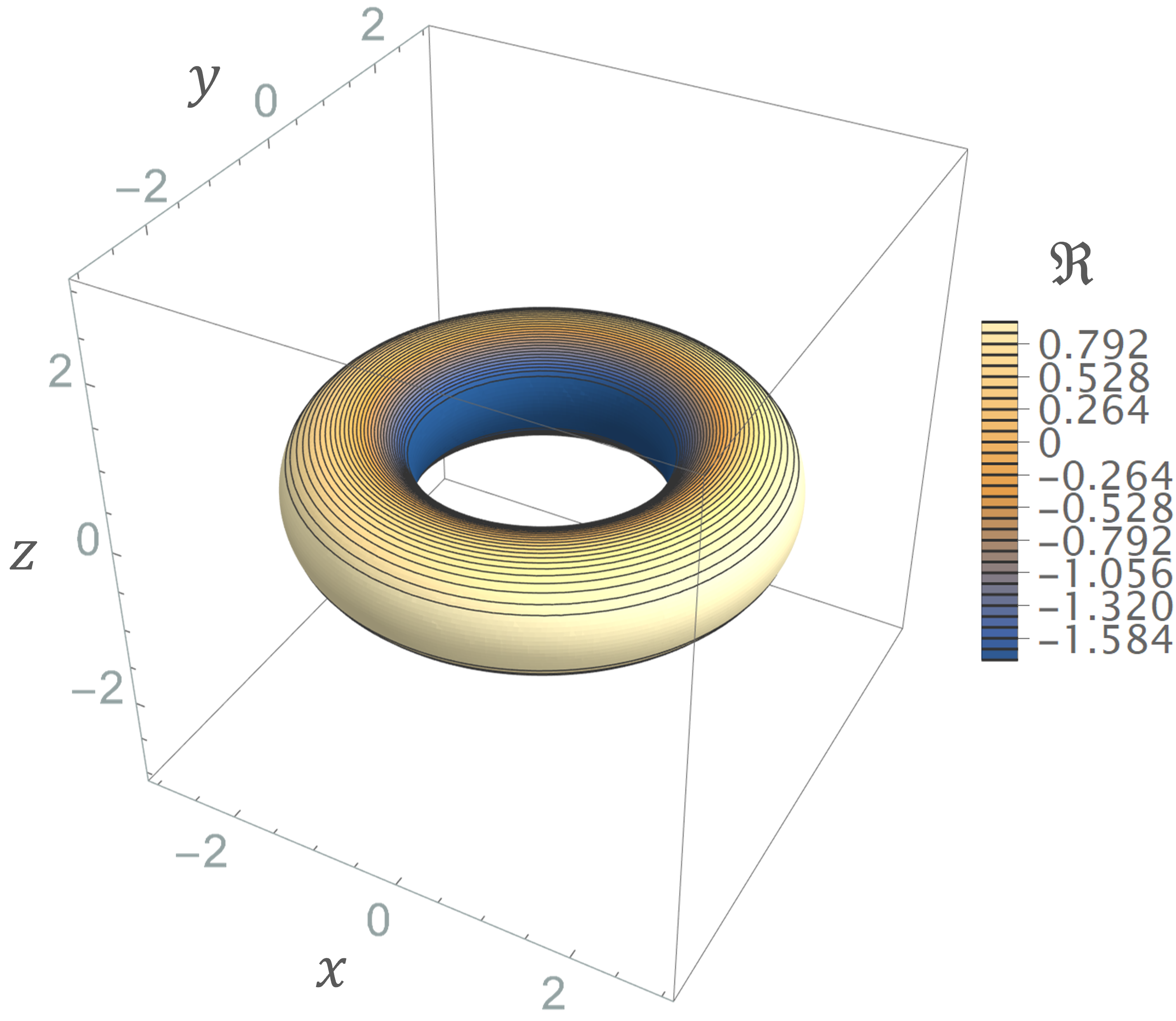}
\caption{\footnotesize Contour plot of the Ricci scalar curvarure \eqref{RiccT} for an axially symmetric torus $T=0.25$ with major radius $r_0=2$.}
\label{fig2}
\end{figure}

We now consider the diffusive equilibrium problem on an axially symmetric torus $\Sigma$ given as a level set of the function
\begin{equation}
T=\frac{1}{2}\left[\lr{r-r_0}^2+z^2\right].
\end{equation}
Here $r_0$ is a positive real constant representing the major radius of the torus, and $r$ the radial coordinate of a cylindrical coordinate system. A convenient coordinate set is $\lr{\mu,\nu,\zeta}=\lr{\phi,z,T}$. The nontrivial components of the corresponding contravariant metric tensor are given by
\begin{equation}
g^{\phi\phi}=\frac{1}{r^2},~~~~g^{zz}=1,~~~~g^{zT}=z,~~~~g^{TT}=2T.
\end{equation}
It follows that the non-vanishing covariant components are
\begin{equation}
g_{\phi\phi}=r^2,~~~~g_{zz}=\frac{2T}{\lr{r-r_0}^2},~~~~g_{zT}=-\frac{z}{\lr{r-r_0}^2},~~~~g_{TT}=\frac{1}{\lr{r-r_0}^2}
\end{equation}
Furthermore, the Jacobian determinant is
\begin{equation}
J=\lr{1-\frac{r_0}{r}}.
\end{equation}
In order to evaluate the Ricci scalar curvature of the surface, we must determine the metric tensor $\gamma_{ij}$ induced on $\Sigma$. 
Since $r=r\lr{z,T}$, the tangent vectors on the torus are
\begin{equation}
\p_{\phi}=-r\sin\phi\nabla x+r\cos\phi\nabla y,~~~~\p_z=\frac{\p r}{\p z}\cos\phi\nabla x+\frac{\p r}{\p z}\sin\phi\nabla y+\nabla z.
\end{equation}
It follows that the nontrivial components of the induced metric tensor are
\begin{equation}
\gamma_{\phi\phi}=r^2,~~~~\gamma_{zz}=1+\lr{\frac{\p r}{\p z}}^2,~~~~\gamma^{\phi\phi}=\frac{1}{r^2},~~~~\gamma^{zz}=\frac{1}{1+\lr{\frac{\p r}{\p z}}^2}.\label{gamma}
\end{equation}
From $\lr{r-r_0}^2=2T-z^2$, it also follows that
\begin{equation}
\frac{\p r}{\p z}=-\frac{z}{r-r_0},~~~~\frac{\p^2 r}{\p z^2}=-\frac{1}{r-r_0}\lr{1+\frac{z^2}{2T-z^2}}
\end{equation}
The Ricci scalar curvature of $\Sigma$ is given by
\begin{equation}
\mf{R}=\frac{2R_{\phi z\phi z}}{\gamma_{\phi\phi}\gamma_{zz}-\gamma_{\phi z}}=\frac{2}{r^2\lr{1+\frac{z^2}{\lr{r-r_0}^2}}}R_{\phi z\phi z},
\end{equation}
where $R_{ijk\ell}$ denotes the Riemannian curvature tensor. 
Evaluating the component $R_{\phi z\phi z}$ gives
\begin{equation}
R_{\phi z\phi z}=\gamma_{\phi m}R^{m}_{z\phi z}=r^2R^{\phi}_{z\phi z}=r^2\lr{\frac{\p\Gamma_{zz}^{\phi}}{\p\phi}-\frac{\p\Gamma_{\phi z}^{\phi}}{\p z}+\Gamma^{\phi}_{\phi m}\Gamma^{m}_{zz}-\Gamma^{\phi}_{zm}\Gamma^{m}_{\phi z}}.
\end{equation}
One can verify that
\begin{equation}
\Gamma_{\phi\phi}^{\phi}=0,~~~~\Gamma_{\phi z}^{\phi}=\frac{1}{r}\frac{\p r}{\p z},~~~~\Gamma_{zz}^{\phi}=0,~~~~\Gamma_{zz}^z=\frac{\frac{\p r}{\p z}\frac{\p^2 r}{\p z^2}}{1+\lr{\frac{\p r}{\p z}}^2}.
\end{equation}
Then,
\begin{equation}
\mf{R}=\frac{2}{r\lr{r-r_0}\left[1+\frac{z^2}{\lr{r-r_0}^2}\right]}.\label{RiccT}
\end{equation}
Notice that this expression reduces to the Ricci scalar curvature of the sphere in the limit $r_0\rightarrow 0$. 
A contour plot of the Ricci scalar curvature for the axially symmetric torus $T$ is shown in figure \ref{fig2}. 

The diffusive equilibrium equation \eqref{DE2} can now be written as
\begin{equation}
\frac{\p^2 w}{\p\phi^2}+\frac{r\lr{r-r_0}}{2T}\frac{\p}{\p z}\left[r\lr{r-r_0}\frac{\p w}{\p z}\right]=-\frac{r\lr{r-r_0}}{T}w.\label{DE3}
\end{equation}
A physically consistent solution should be periodic in the toroidal and poloidal angles $\phi$ and $\vartheta$, with the poloidal angle $\vartheta$ defined by
\begin{equation}
\vartheta=\arctan\lr{\frac{z}{r-r_0}}.
\end{equation}
Performing the change of variables $\lr{\phi,z,T}\rightarrow\lr{\phi,\vartheta,T}$, equation \eqref{DE3} reads as
\begin{equation}
\frac{\p^2 w}{\p\phi^2}+\frac{r}{2T}\frac{\p}{\p\vartheta}\left(
r\frac{\p w}{\p\vartheta}
\right)=-\frac{r\lr{r-r_0}}{T}w.
\end{equation}
The $\phi$-dependence of the equation can be removed by separation of variables, $w=\Phi\lr{\phi}Q\lr{\vartheta,T}$. We have
\begin{subequations}\label{TDE}
\begin{align}
\frac{\p^2\Phi}{\p\phi^2}=&-m^2\Phi,~~~~m\in\mathbb{Z}\\
\frac{\p}{\p\vartheta}\lr{r\frac{\p Q}{\p\vartheta}}=&2\left(
\frac{m^2T}{r}-r+r_0\right)Q.
\end{align}
\end{subequations}
The first equation gives
\begin{equation}
\Phi=A_m\sin\lr{m\phi}+B_m\cos\lr{m\phi},~~~~A_m,B_m\in\mathbb{R}.
\end{equation}
Next, observing that $\lr{r-r_0}^2=2T\cos^2\theta$ and  $z^2=2T\sin^2\vartheta$, the second equation in \eqref{TDE} becomes
\begin{equation}
\lr{\alpha+\cos\vartheta}\frac{\p^2 Q}{\p\vartheta^2}-\sin\vartheta\frac{\p Q}{\p\vartheta}=\lr{\frac{m^2 }{\alpha+\cos\vartheta}-
2\cos\vartheta}Q,\label{TEqX}
\end{equation}
where we introduced the aspect ratio $\alpha^{-1}=\sqrt{2T}/r_0$. 
On a given toroidal surface $T$, this equation must be solved for $Q$ with periodic boundary conditions in $\vartheta$. 
 By noting that the spherical angle $\theta$ is related to the poloidal angle $\vartheta$ according to
\begin{equation}
\sin^2\vartheta=\left(1+2\alpha \cos\vartheta+\alpha^2\right)\cos^2\theta,
\end{equation}
it can be verified that in the spherical limit $r_0=0$ equation \eqref{TEqX} gives the solution \eqref{omS1}.
Interestingly, the vector field $\p_{\phi}$ 
ceases to be a solution of \eqref{TEqX} whenever $r_0\neq 0$ despite $\p_{\phi}$ being a Killing field of the metric induced on the torus \eqref{gamma} (all components $\gamma_{ij}$ are independent of the angle $\phi$). This is because for a
general surface $\Sigma$ the diffusion operator $\mf{D}_{\Sigma}-\mf{R}/\sqrt{g^{\zeta\zeta}}$ and
the Laplacian $\Delta$ restricted to $\Sigma$ do not coincide, and therefore exhibit a different kernel structure. 
Let us look for nontrivial solutions $Q\neq 0$ of \eqref{TEqX} for $r_0\neq 0$. 
For simplicity, we consider the limiting case $\alpha>>1$, which corresponds to a `thin' torus (small aspect ratio). 
At first order in $\alpha^{-1}$, equation \eqref{TEqX} can be written as
\begin{equation}
\lr{\alpha+\cos\vartheta}\frac{\p^2 Q}{\p\vartheta^2}-\sin\vartheta\frac{\p Q}{\p\vartheta}+2\cos\vartheta\,Q=0.\label{Q}
\end{equation}
Observe that $m$ does not appear in the equation, implying that any $m$ is acceptable. Furthermore, this equation is equivalent to what one gets by setting $m=0$ in \eqref{TEqX}. If $Q\neq 0$ the freedom in the choice of $m$ however conflicts with the physical intuition that  diffusive equilibria should not exhibit arbitrary $\phi$ dependent structures due to the axial symmetry of the problem. We therefore expect to find $Q=0$.  
Since the unkown $Q$ must be periodic in the poloidal angle $\vartheta$, we can expand it in Fourier series, $Q=\sum_{k=-\infty}^{\infty}c_k e^{{\rm i}k \vartheta}$. Then,  equation \eqref{Q} becomes a system of equations for the coefficients $c_k$, 
\begin{equation}
-2\alpha k^2 c_k+c_{k-1}\lr{2+k-k^2}+c_{k+1}\lr{2-k-k^2}=0.
\end{equation}
It follows that
\begin{equation}
...,~~~~c_{-3}=-2\alpha c_{-2},~~~~c_{-1}=0,~~~~c_0=0,~~~~c_1=0,~~~~c_3=-2\alpha c_2,~~~~c_4=\frac{1}{5}\lr{18\alpha^2-2}c_2,~~~~...,
\end{equation}
and so on. In particular, one can verify that for $\alpha >>1$ the $k$th coefficient scales as
\begin{subequations}
\begin{align}
c_k\sim &\,\, \alpha^{k-2}c_{2},~~~~k>2.\\
c_{k}\sim &\,\,\alpha^{-k-2}c_{-2},~~~~k<-2.
\end{align}
\end{subequations}
Furthermore, each coefficient can be expressed in the form
\begin{subequations}
\begin{align}
c_k=&A_k c_2,~~~~k>2\\
c_{k}=&A_{k} c_{-2},~~~~k<-2,
\end{align}
\end{subequations}
where $A_k\lr{\alpha}$ is a function of $\alpha$.
It therefore follows that when $\alpha>1$ the Fourier coefficients will diverge for a sufficiently large $\abs{k}$ unless $c_2=c_{-2}=0$. 
Hence, the only well-behaved solution is $Q=0$, which implies a trivial diffusive equilibrium $\omega^{\zeta}=0$.

\section{Concluding remarks}

In this paper, we studied the motion of an incompressible fluid on a $2$-dimensional surface with arbitrary topology embedded in $3$-dimensional Euclidean space. 
First, we derived a vorticity equation for the stream function by using a Clebsch parametrization of the velocity field. 
Then, we identified the conserved energy and enstrophy of the system, and 
obtained the associated Hamiltonian structure in terms of a noncanonical Poisson bracket.
The construction of the diffusion operator for the vorticity equation was also discussed. 
Assuming that microscopic interactions responsible for diffusion are not modified by the embedding, 
the standard approach based on Killing fields of the Riemannian metric gives the $\mathbb{R}^3$ Laplacian 
restricted to the surface. This operator contains derivatives of the stream function across the surface, and
prevents the vorticity equation from being $2$-dimensional. 
We then formulated an alternative diffusion operator by demanding 
consistency with the diffusion operator on the $2$-sphere, and by requiring that 
the outcome 
of the diffusion process on a flat surface is a constant vorticity.    
The resulting vorticity equation is $2$-dimensional.
Furthermore, this diffusion operator, which minimizes an energy dissipation functional weighted by the Ricci scalar curvature,
offers an analogy with the Poisson equation for the gravitational potential in general relativity. 
Examples of the corresponding diffusive equilibria on a $2$-sphere and a $2$-torus were also examined.








\section*{Acknowledgment}
The research of NS was partially supported by JSPS KAKENHI Grant No. 21K13851 and No. 17H01177. 



\end{document}